\documentclass[twocolumn]{aastex62}
\usepackage{url}
\usepackage{natbib}

\newcommand{\bq}{\begin{equation}}
\newcommand{\eq}{\end{equation}}

\newcommand{\ffas}{\hbox{$\,.\!\!^{\prime\prime}$}}

\shorttitle{Evolution of the Gas Mass Fraction}
\shortauthors{Wiklind {\it et al}}

\received{}
\revised{}
\accepted{}

\submitjournal{ApJ}

\begin{document}

\title{EVOLUTION OF THE GAS MASS FRACTION OF PROGENITORS TO TODAY'S MASSIVE GALAXIES:\\
ALMA OBSERVATIONS IN THE CANDELS GOODS-S FIELD}

\author{Tommy Wiklind}
\affiliation{Catholic University of America, Department of Physics, Washington, DC 20064, USA; wiklind@cua.edu}

\author{Henry C. Ferguson}
\affiliation{Space Telescope Science Institute, Baltimore, MD 21218, USA}

\author{Yicheng Guo}
\affiliation{Department of Physics and Astronomy, University of Missouri, Columbia, MO 65211, USA}

\author{David C. Koo}
\affiliation{UCO/Lick Observatory, Department of Astronomy and Astrophysics, University of California, Santa Cruz, 95064 CA, USA}

\author{Dale Kocevski}
\affiliation{Department of Physics and Astronomy, Colby College, Waterville, ME 04901, USA}

\author{Bahram Mobasher}
\affiliation{Department of Physics and Astronomy, University of California, Riverside, CA 92521, USA}

\author{Gabriel B. Brammer}
\affiliation{Cosmic Dawn Center, Niels Bohr Institute, Copenhagen, Denmark}

\author{Susan Kassin}
\affiliation{Space Telescope Science Institute, Baltimore, MD 21218, USA}

\author{Anton M. Koekemoer}
\affiliation{Space Telescope Science Institute, Baltimore, MD 21218, USA}

\author{Mauro Giavalisco}
\affiliation{Department of Physics and Astronomy, University of Massachusetts, Amherst, MA 01003, USA}

\author{Casey Papovich}
\affiliation{Department of Physics and Astronomy, Texas A\&M, College Station, TX 77843, USA}

\author{Swara Ravindranath}
\affiliation{Space Telescope Science Institute, Baltimore, MD 21218, USA}

\author{Sandra M. Faber}
\affiliation{UCO/Lick Observatory, Department of Astronomy and Astrophysics, University of California, Santa Cruz, 95064 CA, USA}

\author{Jonathan Freundlich}
\affiliation{Center for Astrophysics and Planetary Science, Racah Institute of Physics, The Hebrew University, Jerusalem 91904, Israel}

\author{Duilia F. de Mello}
\affiliation{Catholic University of America, Department of Physics, Washington, 20064 DC, USA}

\begin{abstract}
We present an ALMA survey of dust continuum emission in a sample of 70 galaxies in the redshift range $z=2-5$ selected
from the CANDELS GOODS-S field. Multi-Epoch Abundance Matching (MEAM) is used to define potential progenitors of a
$z = 0$ galaxy of stellar mass $1.5 \times 10^{11} M_{\odot}$. Gas masses are derived from the 850$\mu$m luminosity.
Ancillary data from the CANDELS GOODS-S survey are used to derive the gas mass fractions. The results at $z\lesssim3$
are mostly in accord with expectations: The detection rates are 75\% for the $z=2$ redshift bin, 50\% for the $z=3$ bin and
0\% for $z\gtrsim4$.  The average gas mass fraction for the detected $z=2$ galaxies is $f_{\rm gas} = 0.55 \pm 0.12$ and
$f_{\rm gas} = 0.62 \pm 0.15$ for the $z=3$ sample. This agrees with expectations for galaxies on the star-forming main
sequence, and shows that gas fractions have decreased at a roughly constant rate from $z=3$ to $z=0$. Stacked images
of the galaxies not detected with ALMA give upper limits to $f_{\rm gas}$ of $<0.08$ and $<0.15$, for the $z=2$ and $z=3$
redshift bins. None of our galaxies in the $z=4$ and $z=5$ sample are detected and the upper limit from stacked images,
corrected for low metallicity, is $f_{\rm gas}<0.66$. We do not think that lower gas-phase metallicities can entirely explain
the lower dust luminosities. We briefly consider the possibility of accretion of very low-metallicity gas to explain the absence
of detectable dust emission in our galaxies at $z\gtrsim4$.
\end{abstract}

\keywords{cosmology: observations --- galaxies: formation --- galaxies: high redshift --
galaxies: photometry --- galaxies: evolution}

\section{Introduction}\label{sec:intro}

A fundamental process by which galaxies acquire mass is by accreting gas and converting it into stars. This process
is especially important during the epoch extending from the reionization era to the peak of star formation activity at
$z\sim2$. The accretion process and how it is related to the star formation history of these early galaxies is, however,
poorly understood from an observational point of view. We know that the gas accretion rate increases with increasing
redshift, and at $z\gtrsim4$, it may exceed the star formation rate, leading to galaxies whose baryonic content are
dominated by interstellar gas (e.g. Papovich et al. 2011). At lower redshifts, the star formation rate eventually exceeds
the gas accretion rate, leading to a decreasing gas mass fraction. In addition to providing fuel for the star formation activity,
the interstellar gas also fuels activity in the galactic nuclei. The star formation and AGN activities are coeval and follows
the same strong evolution with redshift, showing a broad maximum around $z\sim2$ (e.g. Madau \& Dickinson 2014).
The gas mass fraction in massive galaxies in the local universe, like the Milky Way, is typically $\lesssim$10\% and the
star formation activity has fallen to a level last seen during the end of the reionization epoch at $z$$\sim$7 (e.g.
Madau \& Dickinson 2014).

Despite the importance of gas accretion and the role of the interstellar gas during the formation of galaxies and their
subsequent evolution, almost all of our knowledge about galaxies at high redshift is based on observations of stellar
light. A tight correlation between star formation activity and stellar mass exists at the epoch of maximum star formation
activity, and extends to earlier epochs (Daddi et al. 2007; Noeske et al. 2007; Magdis et al. 2010; Rodighiero et al.
2010, 2011). This correlation evolves smoothly with redshift (e.g. Madau \& Dickinson 2014 and references therein)
and is seen as an indication that the star formation history of most galaxies is driven by an interplay between gas
accretion, star formation and feedback processes. Hence, observations of the stellar component only offers one
piece of an intricate puzzle.

\medskip

Our lack of knowledge about the interstellar medium (ISM) in high redshift galaxies stems from the difficulties
associated with observing the gaseous component. Studies of the warm ionized gas through near-infrared
spectroscopy of rest-frame optical lines provide some information, but in terms of mass this gas component
only represents a small fraction of the total amount of interstellar gas. The cold and dense ISM, the most massive
ISM component and the one directly involved in the formation of new stars, makes its presence known indirectly
through extinction of rest-frame ultraviolet (UV) and optical light but is difficult to observe directly. In the local
universe the atomic and molecular ISM components are usually studied using radio techniques. However, due
to radio interference, the 21\,cm HI line can only be observed in nearby galaxies. Observations of rotational
lines of CO, tracing the more abundant molecular hydrogen, can be observed in high redshift galaxies (e.g.
Carilli \& Walter 2013; Combes 2018). While observation of CO in galaxies at the highest redshifts is possible,
it has mainly been of highly special systems such as massive star forming submillimeter galaxies (SMGs;
Capak et al. 2011; Walter et al. 2012; Riechers et al. 2013; Vieira et al. 2013; Zavala et al. 2018), and AGN
host galaxies, the latter which has been observed at $z\gtrsim7$ (Venemans 2012; 2017). A growing number
of CO observations of main sequence star forming galaxies at redshifts $z\sim1.5-2.5$ are beginning to provide
insights in the molecular ISM of more normal galaxies at high redshift (see Carilli \& Walter 2013; Combes 2018),
but these observations come at a high cost in telescope time, even with a facility like the Atacama Large
Millimeter/submm Array (ALMA). 

\medskip

A faster, albeit less direct, method to estimate the gas mass in high redshift galaxies is to use a single flux density
measurement of optically thin dust emission. This approach uses dust continuum emission as a proxy for the dense
interstellar gas. The optically thin Rayleigh-Jeans part of the dust continuum emission is proportional to the dust
mass, with only a linear dependence on the dust temperature (Scoville et al. 2014). The interstellar gas mass can
then be derived if the gas-to-dust mass ratio is known. Scoville et al. (2014) derived an empirical relation
between the specific luminosity of dust continuum emission at a wavelength of 850$\mu$m and the gas mass
determined through other means (mainly CO observations) for a sample of local Ultraluminous Far-Infrared
Galaxies (ULIRGs) and massive star forming galaxies at $z\sim2$. The ratio $L_{850\mu m}/M_{\mathrm{gas}}$
has a surprisingly small scatter and can thus be used to measure the interstellar gas mass from the dust
continuum emission. This provides a fast and accurate method to derive the gas mass in distant galaxies at
any redshift, and allows studies of the dense and cold interstellar gas at high redshift of statistically meaningful
samples. A caveat with using the dust continuum emission, as well as the CO line, for measuring the gas mass,
is its dependence on the gas-phase metallicity. This is not a problem for galaxies with close to solar metallicity,
but the gas-to-dust mass ratio has a linear dependence on the metallicity (Draine et al. 2007) and care has to
be exercised when observing galaxies of lower stellar mass, possibly with significantly lower than solar metallicity.

\medskip

There have been several recent studies of the gas mass fraction in galaxies at high redshifts. Observations
of CO line emission of galaxies at $z\sim1-2$ shows that the gas mass fraction is increasing dramatically
compared with that found in the local universe (Daddi et al. 2010; Geach et al. 2011; Magnelli et al. 2012;
Tacconi et al. 2013, 2018; Papovich et al. 2016; Freundlich et al. 2019). Using  dust continuum emission as
a proxy for the interstellar gas mass has allowed these studies to be pushed to $z\sim3-4$ (Scoville et al. 2014,
2016; Schinnerer et al. 2016), showing a further increase in the gas mass fraction. These results provide an overall
picture of the growth of the gas mass fraction with look-back time which is largely consistent with expectations from
theory. The samples, however, are selected using different criteria, and do not necessarily reflect an evolutionary
scenario of the same galaxies seen at different epochs, making a comparison across cosmic time difficult.

Observations of dust continuum emission now extends to some of the highest redshifts. For instance, Tamura
et al. (2018) report of thermal dust emission from a $z$=8.31 galaxy. This galaxy, which is also detected in
the [OIII]\,88$\mu$m fine-structure line, is gravitationally lensed by the foreground cluster MACS\,J0416.1-2403.
The de-lensed dust mass is estimated to be $4\times10^6$\,M$_{\odot}$ with a stellar mass of $5\times10^{9}$\,M$_{\odot}$,
showing that metal enhancement has occurred in some very young galaxies in the reionization epoch. It is,
however, not yet possible to assess the gas mass fraction at these very high redshifts as no well-defined samples
have yet been compiled. 

\medskip

As observations probe higher redshifts it becomes increasingly important to define samples that reflect
the true evolution of galaxy properties. Galaxies need to be selected in a way that connects the properties of a
given $z$=0 galaxy, the descendant, with that of its high redshift progenitors, using already known characteristics
of the galaxies such as stellar mass. Several studies have defined galaxy samples across different redshifts based on
constant co-moving number densities (Papovich et al. 2011; Brammer et al. 2011; van Dokkum et al. 2013; Patel et al.
2013; Leja et al. 2013). This provide samples which are better suited for studies of how galaxy characteristics evolve
with cosmic time, than a simple stellar mass cut-off or luminosity selection can do. Another method is to select galaxies at
different redshifts based on their ranking, for instance, by selecting the most massive galaxies in equal co-moving volumes
at different redshifts. Neither of these methods, however, fully account for the effects of galaxy evolution (e.g. Leja et al. 2013),
basically by ignoring scatter and the stochasticity in the mass accretion history of the progenitors (Behroozi et al. 2013).
An alternative selection strategy is to use abundance matching, or more precisely, Multi-Epoch Abundance Matching
(MEAM; Moster et al. 2013; Behroozi et al. 2013). The MEAM method provides a parametrized relation between halo
mass, as derived from numerical simulations, and stellar mass, based on observations of the stellar mass function,
as a function of redshift. This method connects the growth of central dark matter halos with the growth and evolution of the
stellar content and allows us to define the most likely progenitors of a given $z$=0 galaxy while taking the stochastic
nature of galaxy growth into account. While it does not account for all of the baryonic content of galaxies, only the
stellar component, it does provide a recipe for selecting galaxies connected across cosmic time in an evolutionary
sequence.

\medskip

In this paper we present a study of the gas mass fraction in a sample of 70 galaxies selected using
to the MEAM method (Sect.~\ref{sec:meam}) across the redshift range $z=2-5$. {\it The selection is based
only on stellar mass and redshift and is completely unbiased with regards to luminosity, morphology or star
formation activity}. Our sample contains the most likely progenitors of a $z$=0 galaxy of stellar mass
$\log{(M_{*}/M_{\odot})} = 11.2$. The gas mass is derived from optically thin dust continuum emission,
using the conversion between 850$\mu$m specific luminosity and interstellar gas mass derived
by Scoville et al. (2014, 2016). We correct the observed fluxes for the effects of the Cosmic Microwave
Background as well as discuss the impact of low gas-phase metallicity for the galaxies with the lowest stellar
mass in our sample.

\medskip

The paper is structured as follows: Sect.~\ref{sec:sample} presents the method used for defining the
sample of galaxies and its application on the CANDELS GOODS-S data; Sect.~\ref{sec:data} presents
our ALMA observations, data reduction and the corresponding ancillary CANDELS data; Sect.~\ref{sec:results}
presents our results; in Sect.~\ref{sec:discussion} we discuss the results in more detail and explore the impact
of a low gas-phase metallicity. Sect.~\ref{sec:summary} presents a summary of our conclusions. For consistency
with previous CANDELS publications we adopt the following cosmology: $H_{0} = 70$\,km\,s$^{-1}$\,Mpc$^{-1}$,
$\Omega_{\mathrm{M}} = 0.3$ and $\Omega_{\Lambda} = 0.7$.

\section{Sample Selection}\label{sec:sample}

\subsection{The MEAM selection}\label{sec:meam}

A central theme with the study presented in this paper is to compare properties of galaxies at different redshifts,
connected through an evolutionary sequence. To accomplish this we identify the likely high redshift progenitors of
local galaxies of a given stellar mass using the Multi-Epoch Abundance Matching (MEAM) method. The resulting
sample is thus selected based only on stellar mass and redshift, without considering other parameters. Such a
sample is well suited for studying the evolution of properties such as, but not limited to, the gas mass fraction.

\medskip

The method of abundance matching combines results from numerical dark matter simulations with observed
properties of galaxies (Marinoni et al. 2002; Behroozi et al. 2010; Moster et al. 2010). In particular the
multi-epoch abundance matching method developed by Behroozi et al. (2013) and Moster et al. (2013) provides
an approach to determine the relationship between the stellar masses of galaxies and the masses of their host
dark matter halos, while taking the stochastic nature of the merger process as well as observational uncertainties
into account. Instead of attempting to model the complicated baryonic physics associated with star and galaxy
formation, including feedback processes, the abundance matching methods populate dark matter halos, obtained
from numerical simulations, with galaxies using a parametrized empirical model with adjustable parameters. The
only observational input to these methods is the observed stellar mass function at a given redshift. The stellar-to-halo
mass relation (SHM) can then be determined. The MEAM method assumes a functional form for the SHM relation
(see Eq. 2 in Moster et al. 2013), populates the halos with galaxies and computes a model stellar mass function.
The parameters of the SHM are then adjusted until the observed stellar mass function is reproduced. The redshift
evolution of this relation is driven by gas infall, star formation, merging satellite galaxies and stellar mass loss.

\medskip

Using the MEAM method we define a region in $M_{*}$-$z$ space where progenitors of a z=0 galaxy of a given stellar mass are most
likely to be located. We incorporate the statistical scatter in the merger tree from the Bolshoi simulation (Behroozi et al. 2013) and
the $M_{*}(M_{\mathrm{h}},z)$ relation derived from the Millennium simulation (Moster et al. 2013). The result is shown in
Fig.~\ref{fig:slices}, where the different slices correspond to five galaxies of different stellar masses at $z$=0. The selection
regions overlap at high redshift, representing the stochastic nature of the evolution of the merger trees for individual galaxies.
From this $M_{*}(M_{\mathrm{h}},z)$ selection we  define progenitor galaxies to a $z$=0 galaxy of $\log{(M_{*}/M_{\odot})}
= 11.2$, located at redshifts up to $z\sim5$ (shown as the purple area in Figure~\ref{fig:slices}). In order to avoid selecting
galaxies overlapping with lower mass $z$=0 descendants, we sampled galaxies in the upper half of the purple area shown
in Figure~\ref{fig:slices}.  It is worth noting that at all redshifts, the selected galaxies traces the upper branch of the stellar
mass function and thus minimizes the overlap from galaxies with higher stellar masses.  {\it No selection criteria other than
the stellar mass and redshift are used in defining the sample.}

\subsection{The CANDELS GOODS-S sample}\label{sec:candelssample}

The galaxies for our ALMA observations are taken from the CANDELS\footnote{CANDELS: Cosmic Assembly Near-infrared
Deep Extragalactic Legacy Survey; GOODS-S: Great Observatories Origin Deep Survey - South} GOODS-S
survey (Grogin et al. 2011; Koekemoer et al. 2011). We use the CANDELS GOODS-S H-band selected catalog, which
contains 34,930 galaxies (Guo et al. 2013). All galaxies in the CANDELS catalog have photometric or spectroscopic redshifts.
Stellar masses, and other properties have been derived using multiple spectral energy distribution (SED) models (e.g. Santini
et al. 2015). See Section~\ref{sec:candels} for additional details about the CANDELS data.

\medskip

From this catalog we randomly select 70 galaxies in four redshift bins centered on z=2, 3, 4 and 5\footnote{More specifically:
$1.6 < z \leq 2.6$; $2.6 < z \leq 3.6$; $3.6 < z \leq 4.6$ and $4.6 < z \leq 5.8$.}. The redshift bins are referred to as the $z$=2,
3, 4 and 5 samples, respectively. In practice, the selection of galaxies constitutes a continuum from $z\sim1.6-5.8$. The
properties of the 70 galaxies in our sample are listed in Table~\ref{tab:almadata} and Table~\ref{tab:candelsdata}. The
selection is constrained by requiring the galaxies in each bin to have stellar masses within the limits defined through the
MEAM method (Sect.~\ref{sec:meam}). This means that the selected galaxies are likely progenitors of a $z$=0 galaxy of
$\log(M_{*}/M_{\odot})=$11.2 (the purple region in Figure.~\ref{fig:slices}). Each redshift bin contains 20 galaxies, except
the $z$=5 sample where we only have 10 galaxies. The reason for reducing the sample size for the highest redshift bin
is two-fold. The MEAM method of Moster et al. (2013) and Behroozi et al. (2013) is defined up to $z\approx4$, and hence,
the extension to $z\approx5$ is an extrapolation and its validity decreases with increasing redshift. A second reason is
that we are selecting progressively less massive galaxies at higher redshift and need to compensate with longer integrations
to achieve a lower rms noise. As long as the observations probe the rest-frame Rayleigh-Jeans part of the dust SED, the
negative K-correction ensures  that the sensitivity for a given FIR luminosity is approximately constant with redshift (e.g.
Blain \& Longair 1996). However, for similar gas mass fractions, a galaxy with lower stellar mass also has a lower gas mass
and, hence, a lower flux density than a galaxy with a larger stellar mass. The ratio of the average stellar mass of our $z$=2
and $z$=4 samples is $\sim$10. This is partially compensated by an increasing gas mass fraction. A gas mass
fraction of $\sim$0.5 at $z$=2 and $\sim$0.8 at $z$=4 corresponds to an increase in $M_{\mathrm{gas}}/M_{*}$
by a factor of $\sim$4. We therefore need to achieve $\sim$40\% lower rms noise for the $z\gtrsim4$ galaxies
compared to the $z\sim2$ galaxies.

\medskip

We emphasize that our galaxy sample is only defined based on stellar mass and redshift.
Hence, this sample is likely to represent an unbiased selection of the same population of galaxies,
seen at different epochs, unlike a pure mass cut-off or luminosity selected sample.

\section{Data and Analysis}\label{sec:data}

\subsection{ALMA observations and data reduction}\label{sec:alma}

The ALMA observations of our sample were done during the Cycle 3 and 4 observing seasons. The $z$=4 and $z$=5 samples,
targeting a total of 30 galaxies, were observed on April 30, 2016, using 41 12-m antennas. The $z$=3 sample, targeting 20
galaxies, was observed during multiple sessions during the period June 18 - December 14, 2016, using 41-50 12-m antennas.
Finally, the $z$=2 sample, targeting 20 galaxies, was observed December 17-28, 2016, using 50 12-m antennas. The $z$=4--5
observations were done using the Band 6 receivers, 230\,GHz ($\lambda$1.3\,mm), while the $z$=2--3 observations were
done using the Band 7 receivers, 345\,GHz ($\lambda$870$\mu$m). The choice of ALMA bands ensures that we observe
the rest-frame dust continuum on the Rayleigh-Jeans part of the dust SED. The four redshift bins sample the SED at rest-frame
$\lambda_{\mathrm{rest}} = 277, 202, 243$ and $217$\,$\mu$m, respectively. The correlator was configured to process a total
bandwidth of 7.5\,GHz, consisting of four 1.875\,GHz-wide spectral windows. Each of the windows has 128 dual polarization
channels. These channels were combined to produce the final continuum image. 

The data consist of 70 individual observations, one for each of the targeted galaxies. Each of the four redshift intervals were
packed into an scheduling block (SB) and executed until the required rms noise had been achieved. The bandpass calibrator
for the $z$=2 SB was the unresolved QSO J0522-3627, for $z$=3 the QSO J023+1636 was used and for $z$=4 and $z$=5
J0334-4008 was used. Ceres and Pallas were primary flux calibrators. Phase calibrators were J0348-2749 and J0324-2918.
The weather conditions varied during the observations but the overall phase solutions have smooth variations across each
SB execution. The array configurations used during the observations were selected to give the requested angular resolution
of about 0\ffas5$\times$0\ffas5. In fact, the reason for spreading out the observations over more than 6 months was to retain
a uniform angular resolution, thereby achieving a uniform sensitivity for all the data sets. An interferometer measures a finite
number of angular scales, resulting in a filtering of the spatial frequencies. The reconstructed image therefore contains emission
from a relatively small range of angular scales; from the smallest angular scale, usually just called the angular resolution,
to the largest angular scale. Smooth emission on scales larger than the largest angular scale will be filtered out. Hence, retaining
a uniform angular resolution safeguards against losing emission due to these effects\footnote{See {\em Observing with ALMA --
A Primer}: https://almascience.nrao.edu/documents-and-tools/cycle6/alma-science-primer}.

\medskip

The data were calibrated and reduced using the standard ALMA pipeline. The data was of good quality and met the required angular
resolution of approximately 0\ffas5$\times$0\ffas5. The required 1$\sigma$ rms noise was set to 100$\mu$Jy for the $z$=2 and $z$=3
B6 observations. For the $z$=4 and $z$=5 samples, the B6 rms noise requirements were 50$\mu$Jy and 30$\mu$Jy, respectively
(see Sect.~\ref{sec:candelssample}). All the requirements were met with the ALMA data and are listed in Table~\ref{tab:observations}.
The center wavelengths of the continuum data for the B6 and B7 observations are 1,287$\mu$m and 873$\mu$m, respectively. We will
refer to these as the 1.3mm and 870$\mu$m data, or B6 and B7 data, in the rest of this paper. The full half-power beam width (HPBW)
of the ALMA 12-m antennas, our instantaneous field of view, is 26\ffas5 for the B6 data and 18\ffas0 for the B7 data. The combined area
within the HPBW of all 70 ALMA pointings is 7.4 arcmin$^2$. Several off-center submm sources were detected and will be discussed
in a separate paper. 

\medskip

Continuum images of the 70 ALMA pointings were produced using the CASA\footnote{Common Astronomical Software Application}
task {\small CLEAN}. The data were naturally weighted for maximum sensitivity. In the natural weighting scheme, data are weighted
relative to the number of angular scales observed. It provides the highest sensitivity, but does not maximize the angular resolution.
For the latter, a uniform weighting can be used. It is also possible to combine natural and uniform weighting, so called Briggs weighting
(Briggs 1995). We tried different weighting schemes in the {\small CLEAN} process to find the best trade-off between sensitivity and
angular resolution. However, since almost all of the submm detected sources remain unresolved at the nominal 0\ffas5 angular resolution,
we did the final flux measurements using natural weights. The average observational properties of the four redshifts bin are given in
Table~\ref{tab:observations}. The positional accuracy of the peak of the submm emission is $\lesssim$0\ffas1. In the {\small CLEAN}
process we pixelated the image with a pixel size 0\ffas1$\times$0\ffas1. The actual resolution of the interferometer data, however, is larger
than this. For example, the $z$=3 galaxies have an average restoring beam of 0\ffas56$\times$0\ffas40, with a position angle of  $-86^{\circ}$.
Since the units of the continuum image is Jy/beam, the peak intensity corresponds to the integrated flux for completely unresolved sources.
In almost all cases, however, the derived size of the submm emission is slightly larger than the restoring beam. This shows that while the
emission is not resolved, it has an extent which is non-negligible in comparison with the angular resolution of the ALMA data. In these cases
we derived the integrated flux by fitting a two-dimensional Gaussian to the emission. Of the detected galaxies, we found only one case where
the peak flux was larger than the integrated flux. This corresponds to a case where the submm emission originates in two separate regions,
each of them unresolved by the ALMA data. In this case we used the sum of the peak fluxes of the two components when deriving the gas mass.

\medskip

As we target individual galaxies, the identification of detected sources is easier than in a blind survey. Nevertheless,  in order to avoid false
detections we set a detection limit of 5$\sigma$ for a source to be considered securely detected. Of the 26 detected galaxies (see Sect.~\ref{sec:results})
all but two are detected at more than 5$\sigma$. The two marginally detected sources (CANDELS ID: 2344 and 8433) are detected at 3.4 and
4.8 sigma, respectively. The latter is close to our 5$\sigma$ limit and included among our detected galaxies. Close inspection of the CANDELS
ID 2344 shows that the submm emission is associated with a highly inclined star forming galaxy. We therefore treat this as a detection as well.

\subsection{CANDELS data}\label{sec:candels}

Our sample was defined using CANDELS data from the GOODS-S field (Giavalisco et al. 2004; Grogin et al. 2011; Koekemoer
et al. 2011). The redshifts and stellar masses initially used to define the sample came from the Wiki-Z method (e.g. Wiklind et al. 2014) applied
to the CANDELS GOODS-S data. For the analysis presented here we use the official CANDELS catalogs, where photometric redshifts and
stellar masses are compiled using several different teams applying different SED methods (Guo et al. 2013; Santini et al. 2015). Spectroscopic
redshifts are used whenever reliable data are available. Thirteen of the galaxies galaxies in our sample have spectroscopic redshifts.
The differences between the initial selection and the final catalog values are very small and do not affect the the sample selection or the
results presented in this paper. 

\medskip

The CANDELS GOODS-S catalog contains 34,930 H-band selected sources, covering a total of $\sim$170 arcmin$^2$. 
The multiwavelength catalog includes 18 bands and combines observations from the ERS/WFC3 and CANDELS/WFC3
data in the F098M, F105W, F125W, F140W and F160W filters. It includes UV data from both CTIO/MOSAIC and VLT/VIMOS,
as well as optical GOODS-S and CANDELS data in the F435W, F606W, F775W, F814W and F850LP filters. Infrared data
from VLT/HAWK-I Ks (Fontana et al. 2014) and Spitzer/IRAC 3.6, 4.5, 5.8 and 8.0$\mu$m (Ashby et al. 2013) are also part
of the data set. See Guo et al. (2013) for a summary of the CANDELS GOODS-S UV-to-mid-IR data set and corresponding
survey references. The redshifts and stellar mass estimates used in our analysis are from Santini et al. (2015). In addition to
redshift and stellar mass, all of the galaxies in the CANDELS GOODS-S catalog have estimates of the extinction, age, star
formation history and metallicity (see Table~\ref{tab:candelsdata}). These estimates will be used to characterize the galaxies
in our sample. We also use data on the size and morphology of the galaxies in the CANDELS GOODS-S catalog from van der Wel
et al. 2012. The size estimates are derived using Sextractor and the morphological classification relies on the one-dimensional
S\'{e}rsic index derived using the WFC3/F160W data. Given the redshift range of our sample, the S\'{e}rsic index is derived at
rest-frame wavelengths ranging from 5100{\AA} for our $z$=2 galaxies, to 2700{\AA} for the $z$=5 galaxies.

\medskip

In order to improve the results from the CANDELS survey, a team effort to analyze methods used to derive galaxy properties,
such as photometric redshift, stellar mass and other parameters characterizing the galaxy population, was conducted.
Photometric redshift estimates by eleven teams were presented in Dahlen et al. (2013). It was demonstrated that combining
the photometric redshifts from multiple methods reduces the scatter as well as the outlier fraction in photometric redshift values.
A comprehensive study comparing stellar mass estimates of both real galaxies and mock galaxies drawn from semi-analytical models, was
presented in Mobasher et al. (2015). Here, the results of 10 teams showed that the biases are relatively small and mostly present for young
galaxies with ages $\lesssim$100 Myr. In a similar manner, Santini et al. (2015) analyzed the results from 10 different teams fitting SEDs to
the CANDELS GOODS-S and UDS data (Guo et al. 2013; Galametz et al. 2013). The resulting stellar mass catalog represents the CANDELS
public release for the GOODS-S\footnote{Catalogs for GOODS-S are available at \url{http://candels.ucolick.org/data_access/GOODS-S.html}}
and UDS\footnote{Catalogs for UDS are available at \url{http://candels.ucolick.org/data_access/UDS.html}} fields (Santini et al. 2015). We
adopt the CANDELS GOODS-S public catalog for the redshift and stellar mass values of CANDELS GOODS-S galaxies used in this paper.
Stellar masses are based on a Chabrier IMF.
The public releases also contain the results from each of the participating teams of other parameters obtained from the SED fits. For
these parameters we use median values for dust extinction, $E_{\mathrm{B-V}}$, metallicity, star formation rate (SFR) and specific star
formation rate (sSFR). The CANDELS derived parameter values for the galaxies in our sample are listed in Table~\ref{tab:candelsdata},
together with average values for the different redshift bins.

\section{Results}\label{sec:results}

\subsection{Submm data}\label{sec:submm}

The data obtained with our ALMA observations are listed in Table~\ref{tab:almadata}.
We detect submm emission in 15 out of 20 galaxies in the $z$=2 sample, 11 out of 20 in the $z$=3 sample and none in the $z$=4 and $z$=5
samples, with 20 and 10 galaxies observed, respectively. When detected, the submm emission is in almost all cases observed at $>$5$\sigma$, while
for the non-detected sources, there is no hint of emission.  We construct stacked ALMA images of the non-detected galaxies for each redshift bin.
The number of galaxies in each stack is 5, 9, 20 and 10 for the $z$=2--5 samples, respectively. The rms noise for the stacked $z$=4 and 5
samples is $\sim$8$\mu$Jy and $\sim$35$\mu$Jy for the $z$=2 and 3 samples (see Table~\ref{tab:stacked}). No submm
emission was detected in any of the stacked images. For the $z$=3--5 samples we randomly picked sub-samples of four galaxies
and stacked the submm images. None showed any emission even at the 2$\sigma$ level.
We use the upper limits of submm emission in the stacked images to set upper limits to the gas mass fraction.

\medskip

Conversion of the submm fluxes into estimates of the total molecular gas mass is done using a single flux density measurement of the
dust continuum on the Rayleigh-Jeans part of the dust SED. We use the empirical correlation between the rest-frame 850$\mu$m luminosity,
$L_{850\mu m}$, and the total molecular gas mass, $\alpha_{850\mu m}$, derived by Scoville et al. (2014, 2016):
\begin{eqnarray}\label{eq:alpha}
\alpha_{\nu_{850\mu m}} & = & \frac{L_{850\mu m}}{M_{\mathrm{gas}}} = 6.7\pm1.7 \times 10^{19}\ \mathrm{erg\,s^{-1} Hz^{-1} M_{\odot}^{-1}}
\end{eqnarray}
This correlation was obtained using local and high redshift galaxies with known submm luminosities and known molecular gas masses.
The relation has an approximately linear dependance on dust temperature such that an increase in the dust temperature results in a smaller
gas mass for a fixed flux density. The method was tested on local galaxies with approximately solar metallicity (e.g. Scoville et al. 2014; Groves
et al. 2015), and for high redshift galaxies (Scoville et al. 2016). The high redshift galaxies were selected to have high stellar masses and should
therefore have close to solar metallicity (Savaglio et al. 2005; Erb et al. 2006). 

This approach has recently been explored using high-resolution cosmological simulations from the Feedback in Realistic Environment (FIRE) project
(Hopkins et al. 2014; Liang et al. 2018). The simulations include galaxies at redshifts $z$=2-4, with $M_{\*} \gtrsim 10^{10}$\,M$_{\odot}$, exhibiting
various forms of stellar feedback, and use radiative transfer modeling to estimate the FIR luminosity. The results show a tight correlation between the rest-frame
$L_{850\mu m}$ luminosity and molecular gas mass for a wide range of star formation activity (Liang et al. 2018). This shows that the empirical approach
of using a single flux density measurement to estimate the interstellar gas mass is viable.
Nevertheless, one concern for our sample  is that we target galaxies with stellar masses decreasing with increasing redshift and hence galaxies in
our highest redshift bins may have a different conversion factor between $L_{850\mu m}$ and total gas mass than their lower redshift counterparts.
We will discuss this in more detail in Section~\ref{sec:discussion}.
The gas mass of each galaxy is derived using the conversion factor $\alpha_{850\mu m}$ as defined in Eq.~\ref{eq:alpha}, assuming a dust
temperature $T_{\mathrm{d}}=25$\,K and dust emissivity index $\beta=1.8$ to derive $L_{850\mu m}$ (e.g. Scoville et al. 2016).
A dust temperature of 25\,K is likely to characterize the bulk of the dust mass.  The median dust temperature of the dust in our Milky Way galaxy is
$\sim$18\,K (e.g. Planck Collaboration 2011). Dunne et al. (2011) found that typical dust temperatures for 1867 galaxies in the {\em Herschel-ATLAS}
survey are in the range 17-30\,K.
 
 \medskip
 
The observed submm fluxes at high redshifts need to be corrected for the effects of an increasing temperature of the Cosmic Microwave Background
(CMB); see da Cunha et al. (2013) for a discussion and quantification of this correction. The CMB affects the observed fluxes by providing an additional
heating source for the dust grains as well as an increasingly bright background against which the continuum emission is measured. These two effects
compete against each other, but the net result is that the observed flux density needs to be corrected by a factor $\gtrsim$1. The correction factor
depends on the dust temperature, the observed wavelength and the redshift. For typical galactic dust temperatures, the correction becomes significant
already at $z\sim3-4$. The correction factors used here are listed in Table~\ref{tab:almadata}, for an assumed dust temperature of 25\,K. The correction
increases the flux density values, and hence the estimated gas mass, by $\sim$6\% at $z\sim2$ and by $\sim$30\% at $z\sim5$.
Due to the definition of the gas mass fraction used here, the effect on $f_{\mathrm{gas}}$ is of a lesser magnitude.
For equal stellar and ISM masses, the CMB correction increases $f_{\mathrm{gas}}$ by $\sim$13\% at $z\sim5$ and $\sim$2\% at $z\sim2$. All the
results quoted in this paper and listed in Table~\ref{tab:almadata} have been corrected for the CMB temperature at the corresponding redshifts.

\subsection{Gas mass fraction}\label{sec:gmf}

The gas mass fraction is defined as $f_{\mathrm{gas}} = M_{\mathrm{gas}}/(M_{\mathrm{gas}} + M_{*})$. Upper limits to $M_{\mathrm{gas}}$
and the gas mass fraction $f_{\mathrm{gas}}$ correspond to 3$\sigma$ limits of the flux density. The stellar mass is obtained from the CANDELS
GOODS-S data (Santini et al. 2015; see also Mobasher et al. 2015).

In Figure~\ref{fig:scalinglaws} we plot the gas mass fraction, $f_{\mathrm{gas}}$, as a function of redshift. For the $z$=2-3 redshift bins we plot
both detected galaxies and 3$\sigma$ upper limits of undetected galaxies, as well as the average gas fraction for detected galaxies in the $z$=2
($f_{\mathrm{gas}} = 0.55 \pm 0.17$), and the $z$=3 ($f_{\mathrm{gas}} = 0.62 \pm 0.12$) redshift bins (the errors represent the dispersion
of the mean). These values are significantly higher than the  typical gas mass fraction in large spiral galaxies in the local universe of $\lesssim$0.1.
The corresponding ISM gas masses range from $0.25 - 3.2 \times 10^{11}$\,M$_{\odot}$, with an average of of $1.0 \pm 0.9 \times 10^{11}$\,M$_{\odot}$
for the $z$=2 sample, and $0.7 \pm 0.4 \times 10^{11}$\,M$_{\odot}$ for the $z$=3 sample. In Figure~\ref{fig:scalinglaws} we also show the 3$\sigma$
upper limits of the gas mass fraction for the stacked images of non-detected galaxies in each redshift bin (purple arrows). 
The expected gas mass fraction from three different scaling relations are also shown in Figure~\ref{fig:scalinglaws} (Sargent et al. 2014; Scoville et al. 2017;
Tacconi et al. 2018). The scaling relations have been used to calculate the expected gas mass fraction $f_{\mathrm{gas}}$ for galaxies on the main sequence
(MS) with a stellar mass, $M_{*}(z)$, corresponding to the average of the MEAM selection used for our sample (Sect.~\ref{sec:meam}, Figure~\ref{fig:slices}).
Overall, the measured $f_{\mathrm{gas}}$ values of the detected galaxies exhibit the expected increase with redshift for the $z$=2 and $z$=3 samples.
The stacked upper limits (3$\sigma$ limits) on the other hand, deviate significantly from the scaling laws applicable for galaxies on the main sequence.

\medskip

At any given redshift, our selection of galaxies is based on the galaxy stellar mass with a dispersion given by the MEAM models (Sect.~\ref{sec:meam};
Figure~\ref{fig:slices}). We therefore derive the $\Delta f_{\mathrm{gas}} = (f_{\mathrm{gas}})_{obs} - (f_{\mathrm{gas}})_{model}$ for each of the
26 galaxies with detected submm emission where we use the actual stellar masses, star formation rates and, in the case of the Tacconi et al. (2018)
scaling relations, the effective radius, $R_{\mathrm{e}}$ to calculate $(f_{\mathrm{gas}})_{\mathrm{model}}$. The result is shown in Figure~\ref{fig:scatter_0}.
The average $\Delta f_{\mathrm{gas}}$ for the Sargent et al. (2014) and Tacconi et al. (2018) scaling laws are $0.013 \pm 0.15$ and $0.025 \pm 0.15$,
respectively, showing an overall good correspondence between the observed and expected gas mass fractions for the galaxies with detected submm emission.

\medskip

The fraction of galaxies with quenched star formation decreases with increasing redshift, as well as with decreasing stellar mass. One would
therefore expect that the fraction of galaxies in our sample with detectable submm emission increases with redshift. This is not the case and
instead our detection rate decreases from 75\% at $z$=2, to 50\% at $z$=3 and 0\% for $z\geq4$. Could this be due to a lack of sensitivity
in our ALMA data or due to the random selection of galaxies within each redshift bin and a modest sample size?

Quiescent and star forming galaxies occupy distinct regions of the rest-frame $U-V$ versus $V-J$ color space (e.g. Williams et al. 2009).
While all of the submm detected galaxies in our $z$=2 and $z$=3 samples have $UVJ$ colors defining them as star
forming, 3 out of the 5 undetected galaxies in the $z$=2, and 5 out of the 9 undetected galaxies in the $z$=3 sample, have $UVJ$ colors implying
they are quiescent. Incidentally, none of the galaxies in our $z\gtrsim4$ bins fall in the quiescent region, but the $UVJ$ color scheme is not well
defined for these high redshifts. The regions in color space defining `quiescent' and `star forming' galaxies are binary in nature, although it does
appear to distinguish between disk-like and spheroidal morphological types for $z\lesssim2$ (Patel et al. 2012). A more nuanced view of the
star forming activity can be obtained through the extensive CANDELS photometry and corresponding SED fits. In particular, the SFR and stellar
mass estimates allow us to define the sSFR. In Figure~\ref{fig:ssfr} we plot sSFR versus stellar mass for galaxies in each of our four redshift bins.
Each bin contains all CANDELS GOODS-S galaxies, for that particular redshift interval. We mark those galaxies that fall within our $M_*-z$ MEAM 
selection limits, as well as the  galaxies observed with ALMA. The main sequence sSFR for each redshift bin, and its $\pm$0.5 dex values, are marked
(from Schreiber et al. 2015).
From the figure it is clear that 2 out of 5 undetected galaxies in the $z\sim2$ sample have sSFR values 0.5-0.7 dex below the main sequence, and 6
out of 9 undetected galaxies in the $z\sim3$ sample have sSFRs 0.5-0.8 dex below the main sequence. These galaxies corresponds to the 
`quiescent' galaxies in the $UVJ$ color definition. We also note that the both the $z$=2 and $z$=3 redshift bins have undetected galaxies with
sSFRs corresponding to galaxies on the MS as well as 0.5 dex above the MS.
From Figure~\ref{fig:ssfr} it is clear that there are galaxies within our MEAM selection regions with sSFR values 0.1-0.3\% of the corresponding MS value.
These are likely to be truly quiescent galaxies, but none of these are part of our ALMA sample\footnote{This is by pure chance as we selected galaxies
randomly within each MEAM selection region.}.

\medskip

The galaxies in the Tacconi et al. (2018) sample (cf. Freundlich et al. 2019), are selected based on their SFR and M$_{*}$ properties, i.e. their
sSFR, aiming for a homogeneous coverage of the M$_{*}$--SFR plane above a certain stellar mass. Some of these galaxies do lie in the
quiescent part of the UVJ diagram, despite being relatively close to the MS.
The Tacconi et al. (2018) scaling relations are defined for stellar masses in the range $\log{M_{*}/M_{\odot}} = 9.0-11.8$ and across
$\log{\Delta MS} = -1.3 - 2.2$. All of the galaxies in our sample falls within this mass range, and all of the inferred sSFRs are within $+0.5$ and $-0.8$
dex of the MS. Hence, the Tacconi et al's scaling relations is applicable to our sample. For the undetected $z\sim2$ and $z\sim3$ galaxies, the scaling
law predicts $f_{\mathrm{gas}} = 0.37$ and $0.38$, respectively. This is lower than for the detected galaxies but still higher than the upper limits derived
from the stacked images ($f_{\mathrm{gas}} < 0.08$ and $<0.15$). This will be discussed further in Sect~\ref{sec:stacked}.

\medskip

The SFRs used here are obtained from the multi-team SED fitting of the CANDELS data (Santini et al. 2015) and listed in Table~\ref{tab:candelsdata}.
In order to determine whether our ALMA data is sensitive enough to detect these lower gas mass fractions, we need to relate the SFR to an expected
observed submm flux density. We can convert the SFRs into FIR luminosities using the Kennicutt (1998) relation. Fitting a dust SED to our ALMA data
using a modified black-body curve allows us to use the 3$\sigma$ noise rms to obtain lower limits to the detectable SFR.
However, the FIR luminosity derived from the modified blackbody SED has a strong dust temperature dependence, as well as weaker dependencies
on the dust emissivity index $\beta$ and the critical wavelength distinguishing between optically thick and thin emission (e.g. Scoville et al. 2014 and references
therein). The dust emission therefore needs to be well-sampled by observations in order to provide a useful estimate of $L_{\mathrm{FIR}}$ and
corresponding SFR. This is not the case for our sample. We can, however, obtain a rough estimate of the minimum SFR that would result in detectable submm emission for a given
rms noise level by using a modified blackbody SED and vary the dust temperature and other parameters within reasonable limits. Integrating the
resulting SED gives the $L_{\mathrm{FIR}}$ corresponding to our 3$\sigma$ detection limit.
We vary the dust temperature between 25-35 K, the dust emissivity index $\beta=1.5-1.8$ and the critical wavelength $\lambda_0 = 50-100$\,$\mu$m.
This results in a lower limit to the SFR for individual galaxies in the range 4\,M$_{\odot}$\,yr$^{-1}$ ($T_{d}=25$\,K, $z=5$) to 14\,M$_{\odot}$\,yr$^{-1}$
($T_{d}=35$\,K, $z=2$). Lower SFR values are obtained for the lower values of the dust temperature. With all parameters fixed, the lowest SFR values
are obtained for the highest redshift bins due to the lower rms noise levels of the ALMA data. Assuming a dust temperature of 35\,K, the SFR limits are between
9-14 M$_{\odot}$\,yr$^{-1}$.
Using the 3$\sigma$ rms noise from the stacked images we find that a star formation rate of 1-2 M$_{\odot}$\,yr$^{-1}$ should have resulted in
detectable submm emission. The latter limit assumes $T_{\mathrm{d}}=35$\,K. For a lower dust temperature, the limiting SFR decreases even
further.

Hence, even in the case of a high dust temperature, $T_{\mathrm{d}}\sim35$\,K, all of the $z$=2 galaxies and all except three $z$=3 galaxies
ought to have been detected in our ALMA data. The fact that 35\% of the $z$=2-3 galaxies were not detected means that they either lack the
star formation required to power the dust emission and/or or have low dust content and gas-phase metallicities. For the $z$=4 and $z$=5 galaxies,
almost half of the individual galaxies could remain undetected in our ALMA data if the dust temperature is as high as 35\,K, but the other half
should have produced detectable submm emission. If the dust temperature is 25\,K we should have detected all except $\sim$3-4 galaxies in
the $z$=4--5 samples.

The SFR limits for the stacked galaxies without detected submm emission are in the range 1-2 M$_{\odot}$\,yr$^{-1}$, for $T_{\mathrm{d}}=35$\,K.
The fact that none of them show any hint of submm emission shows that these galaxies either have very low levels of star formation or a low gas-phase
metallicity making the dust signature faint enough to evade detection. We will discuss this further in Sect.~\ref{sec:discussion}.

\subsection{Comparison with CANDELS data}\label{sec:candelsdata}

The availability of CANDELS data for all galaxies in our sample makes it possible to compare the derived gas mass fractions with characteristics
of the galaxies derived from SED fits, such as morphology (characterized through the one-dimensional S\'{e}rsic index), the global dust extinction
($E_{\mathrm{B-V}}$), stellar metallicity and effective radius, among others.  These parameters are derived using SED fits to the UV-to-NIR
photometric data using different fitting algorithms, stellar isochrones, parametrization of the star formation history (SFH) and the inclusion, or
omission, of nebular lines. In order to study, and account for, the impact of these effects, the release of CANDELS catalogs involved the effort
of several teams within the CANDELS collaboration (e.g. Mobasher et al. 2015; Santini et al. 2015).

\medskip

If the non-detected galaxies are quiescent systems, we expect them to have high S\'{e}rsic indices, low $E_{\mathrm{B-V}}$ values as well
as low SFR and sSFR values. These expectations are, for the most part, realized, as shown in Figures~\ref{fig:scatter_1}-\ref{fig:scatter_3}.
In these figures we plot $f_{\mathrm{gas}}$ vs. S\'{e}rsic index, effective radius ($R_{\mathrm{e}}$), SFR, sSFR, extinction ($E_{\mathrm{B-V}}$)
and stellar metallicity. Individual results are shown for the detected $z$=2 and $z$=3 galaxies and for the corresponding averages. For the
undetected galaxies, we show the 3$\sigma$ upper limit to $f_{\mathrm{gas}}$ from the stacked images versus averages for the different
parameters.

\medskip

In Figure~\ref{fig:scatter_1} we plot the gas mass fraction, $f_{\mathrm{gas}}$, versus the S\'{e}rsic index (left panel) and the effective radius,
$r_{\mathrm{e}}$, in kpc (right panel).
The $z$=2 and $z$=3 samples show a clear trend with the detected galaxies having a S\'{e}rsic index indicating a disk-like morphology, while the
undetected galaxies, on average, have a more spheroidal morphology. The $z$=4 and 5 samples fall in between the
detected and undetected $z$=2 and 3 samples. However, the uncertainty of the S\'{e}rsic indices for these high redshift galaxies is substantial.
The right panel of Figure~\ref{fig:scatter_1} shows that the submm detected galaxies are,
on average, a factor two larger than the non-detected galaxies. The dispersion in $r_{\mathrm{e}}$ is quite large but our results for $z$=2-3
are consistent with the results of other studies (e.g. van Dokkum et al. 2008), where the effective radius for quiescent galaxies at $z\sim$2 is
$\sim$0.9\,kpc. Star forming galaxies are typically $\sim$3 times larger (e.g. Straatman et al. 2015).

\medskip

In Figure~\ref{fig:scatter_2}, we compare $f_{\mathrm{gas}}$ with the SFR and sSFR. In this case we use median
values as each redshift bin shows a spread in SFR properties (see Table~\ref{tab:candelsdata}). Not surprisingly, the galaxies without  detectable
submm emission tend to have lower SFRs than those with detectable FIR emission. However, the median SFR of the undetected $z$=2 sample
is still $\sim$50\,M$_{\odot}$\,yr$^{-1}$, while the median SFR for the $z$=2 and 3 samples with submm emission is $\gtrsim$100\,M$_{\odot}$\,yr$^{-1}$.
The $z$=4 and 5 samples have the lowest SFRs, with values $\sim$10-15\,M$_{\odot}$\,yr$^{-1}$. However, the $z$=4 and 5 samples show a
very different behavior for the star formation rate per unit stellar mass (sSFR). In this case, the $z$=4 and 5 galaxies have a normal sSFR (right
panel in Figure~\ref{fig:scatter_2}), while the nondetected $z$=2 and 3 galaxies have low sSFR values. The submm detected
galaxies have median sSFR that would put them close to the main sequence at their respective redshifts. This can also be seen in Figure~\ref{fig:ssfr}. 

\medskip

Finally, in Figure~\ref{fig:scatter_3} we compare the gas mass fraction with dust extinction, characterized through $E_{\mathrm{B-V}}$,  and the
stellar metallicity ($Z/Z_{\odot}$). Not surprisingly, the galaxies with detectable submm emission have higher extinction values than the undetected
galaxies. The lowest extinction values are found for galaxies in the $z$=4 and $z$=5 redshift bins. Again, this is not surprising since our sample is
drawn from an H-band selected catalog. At $z\gtrsim4$, the H-band samples rest-frame UV wavelengths of $\lambda \lesssim 3000$\,{\AA}.
Hence, the H-band selection has an intrinsic bias against dust obscured galaxies at these high redshifts. This will be discussed in Sect.~\ref{sec:discussion}.
There is a clear trend with high gas mass fraction galaxies in the $z$=2 and 3 samples having a higher metallicity than those of the undetected
galaxies, with an average metallicity $Z = 1.3\pm0.5$\,$Z_{\odot}$.
A somewhat surprising result is that the average metallicity of the galaxies in the  $z$=4--5 redshift bins have a metallicity close to solar,
$Z=0.9 \pm0.5$\,Z$_{\odot}$. The metallicity of each galaxy is derived from SED fits and thus represent the metallicity of the stellar component, not the
gas-phase metallicity.
The metallicity derived from SED fits are subject to a degeneracy with stellar age and dust extinction. Hence, the metallicity
derived for a single galaxy is not very accurate, but as an average of a larger sample, the SED based stellar metallicities are indicative of a trend
(see Sec.~\ref{sec:metallicity}).

\medskip

Randomly selected galaxies from the $z$=2 and $z$=3 samples are shown in Figure~\ref{fig:sources_1}. For each redshift bin we show
5 submm detected and 5 non-detected galaxies. The morphological difference between the detected and non-detected galaxies in the 
$z$=2 sample, also seen in their S\'{e}rsic indices, is obvious, but it is less evident in the $z$=3 sample.
In Figure~\ref{fig:sources_2} we show 5 randomly selected galaxies each for the $z$=4 and the $z$=5 samples. These galaxies are less massive
and less luminous, which, combined with the surface dimming, makes their appearance at rest-frame UV and optical wavelengths more difficult
to characterize. 
For each galaxy we show 7\ffas0$\times$7\ffas0 cut-outs in the ACS/F606W, ACS/F850LP, WFC3/F160W and VLT/HAWKI 2.2$\mu$m filters,
as well as the ALMA continuum image. The stretch of the ALMA continuum images are different in each case.
The $z\sim5$ galaxy CANDELS ID \#17427 appears to have a weak and extended emission feature associated with the target galaxy. However,
the SNR is $\lesssim$2 and we do not treat this as a detection. CANDELS ID \#6780 (a non-detected $z\sim3$ galaxy) has a neighbor with dust
continuum but at a photometric redshift of $z$=1.550 and therefore not physically related to the target galaxy and not part of our sample.

\section{Discussion}\label{sec:discussion}

\subsection{The gas mass fraction over cosmic time}

In Figure~\ref{fig:averages} we compare our results of the gas mass fraction with results compiled from the literature.
The $f_{\mathrm{gas}}$ estimates are derived from both CO J=1-0 through J=3-2 line observations (Geach et al. 2011;
Daddi et al. 2010; Magnelli et al. 2012; Tacconi et al. 2013, 2018; Papovich et al. 2016; Freundlich et al. 2019) and from dust continuum emission
(Scoville et al. 2016; Schinnerer et al. 2016; Tacconi et al. 2018; this paper).
We have included data from the PHIBSS2 survey (Tacconi et al. 2018), containing gas mass estimates from both CO line and
dust continuum for 1444 star forming galaxies at redshifts
$z\sim0.1-4$. Here we include galaxies up to $z\sim2.5$. We also include CO data from Freundlich et al. (2019), covering the redshift
range $z=0.5-0.8$.
The estimates of the gas mass fraction presented in Figure~\ref{fig:averages}  are based on galaxies selected using different criteria,
mainly stellar mass and star formation activity, and do not necessarily constitute an evolutionary sequence in the same sense as our
selection is meant to do. However, all the observations at $z\lesssim3$ select relatively massive galaxies, $\log{(M_{*}/M_{\odot})}
\gtrsim 10.2$ and it is only for the highest redshifts that our sample selection differ by targeting lower mass galaxies. The errors
shown in Figure~\ref{fig:averages} are dispersion of the mean and do not reflect uncertainties associated with the different estimates
of mass, conversion factors or other possible systematic effects.

Schinnerer et al. (2016) selected a sample of massive, star forming galaxies in the COSMOS field with redshifts $z\sim3.2$ and found
an average $f_{\mathrm{gas}} = 0.62 \pm 0.10$, which is identical to the gas mass fraction derived from our $z$=3 sample, $f_{\mathrm{gas}} =
0.62 \pm 0.12$. The stellar masses of the galaxies making up the Schinnerer et al. (2016) sample is almost the same as for our $z$=3
sample, $\log{(M_{*}/M_{\odot})} = 10.7$ and $\log{(M_{*}/M_{\odot})} = 10.6$, respectively. Scoville et al. (2016) used dust continuum
emission to estimate the gas mass fraction in three redshift bins, $z$$\sim$1.2, 2.3 and 4.4. Their result for $z\sim2.2$ gives
$f_{\mathrm{gas}} = 0.40 \pm 0.16$ for galaxies with an average stellar mass $\log{(M_{*}/M_{\odot})} = 11.0$. Our result for the
$z$=2 sample suggests a higher gas mass fraction at this redshift, with $f_{\mathrm{gas}} = 0.55 \pm 0.17$, for galaxies that on
average are 1.7 times less massive. Tacconi et al. (2013) used the CO, J=3-2 emission line to estimate the gas mass fraction in a
sample of galaxies at $z$=2.2. Using a Galactic value for the CO-to-H$_2$ conversion factor they derive an average gas mass
fraction of $\sim$0.47. The average stellar mass of their sample of $\log{(M_{*}/M_{\odot})} = 10.7$.

\medskip

Figure~\ref{fig:averages} clearly shows that the gas mass fraction
increases dramatically with redshift. The curve shown in the figure depicts the expected gas mass fraction from the scaling relation
of Sargent et al. (2014) for galaxy stellar masses following our selection method. The data shows a steeper increase in the gas mass
fraction for $z\lesssim2$ than indicated by the scaling relation. The scaling relation from Scoville et al. (2018) depicts a steeper rise in
the gas mass fraction for this redshift range, but still falls below the observed values for $z\lesssim2$ (see Figure~\ref{fig:scalinglaws}).
The scaling laws used here are based on MS galaxies with stellar masses given by our MEAM selection and do not take individual star
formation parameters into account.

\medskip

If we plot the observed gas mass fractions as a function of cosmic time rather than redshift, they can be fitted by a linear relation:
$f_{\mathrm{gas}} = (0.075\pm0.055) + (0.043\pm0.007)\,t_{\mathrm{Gyr}}$. The fit is made over the range $z$=3 to $z$=0 and
include all the corresponding data points shown in Figure~\ref{fig:averages}. This result shows that the gas mass fraction for galaxies
in the mass range $\log{M_{*}/M_{\odot}}\approx 10.7-11.2$, on average, decreases at a constant rate of $0.043\pm0.007$ Gyr$^{-1}$
over a time interval of $\sim$11.5\,Gyr. The fit is shown, as a function of redshift, in the right panel of Figure~\ref{fig:averages}
(thick black line). 

\medskip

Galaxies with a larger stellar mass are believed to have a lower gas mass fraction. The observed down-turn in the gas
mass fraction for $z\gtrsim$4, as seen in Figure~\ref{fig:averages}, is therefore most likely attributed to the fact that the $z\sim4$ sample
of Scoville et al. (2016) contains galaxies that have an average stellar mass 15 times higher than those for which the scaling laws shown
in Figure~\ref{fig:averages} are based. This is illustrated in Figure~\ref{fig:scoville} were we plot the expected gas mass fraction as a function
of stellar mass for a fixed redshift of $z$=4.4. In fact, a tendency for this is seen in the Scoville et al. (2016) data as it covers a range of stellar
masses. The same trend is also reflected in the scaling relations (Sargent et al. 2014; Scoville et al. 2017; Tacconi et al. 2018).

\medskip

Hence, several studies, using both dust continuum emission and CO line emission as estimates of the interstellar gas mass, agree that
the gas mass fraction at redshifts $z\sim2$ and $z\sim3$ are $\sim$0.5 and $\sim$0.6, respectively. There may be a dependence
on the galaxy stellar mass in the sense that for a given redshift, less massive galaxies have a higher gas mass fraction than more massive
ones. These gas mass fractions should be compared with the situation in the local universe, where massive galaxies, like our Milky Way,
typically have $f_{\mathrm{gas}}\lesssim0.1$, while dwarf galaxies can have much higher gas mass fractions (e.g. Schombert et al. 2001;
Bergum et al. 2008).
Taking the disparity of the sample selection into account, as well as the disparity in the techniques used to infer the interstellar
gas masses and the associated potential systematic effects and biases, the consistency in the dramatic rise of the gas mass
fraction with redshift is quite remarkable and reflects the profound importance that the interstellar gas has on regulating
the star formation history of galaxies as they evolve with cosmic time.
Adopting a fiducial gas mass fraction of 0.08 at $z$=0, these results show that the gas mass fraction grows as
$f_{\mathrm{gas}} \propto (1 + z)^{1.5}$ over $0 \leq z \leq 3$, and even steeper if we only consider the $0 \leq z \leq 2$
redshift range.

\subsection{Non-detections of dust continuum from stacked images}\label{sec:stacked}

One intriguing aspect of the gas mass fraction estimates for galaxies in our sample is the result for stacked
images of galaxies without detectable dust continuum emission. In Figure~\ref{fig:scalinglaws} these upper limits are marked
by purple arrows and represent upper limits to the gas fraction with a 3$\sigma$ upper limit of the gas mass. In Figure~\ref{fig:averages}
the $z$=4 and $z$=5 upper limits are marked as red arrows. These limits are significantly lower than the gas fractions of the
detected galaxies at $z$=2-3 as well as the expected gas mass fraction derived from scaling laws. The upper limits to the gas
fractions are listed in Table~\ref{tab:stacked}.

\medskip

As discussed in Sect~\ref{sec:gmf}, about half of the non-detected galaxies at $z\sim2$ and $z\sim3$ have sSFRs lower than
those that are submm detected (Figure~\ref{fig:ssfr}). The sSFR of the non-detected galaxies are within a factor $\sim$0.8 dex of
the sSFR of main sequence at the corresponding redshift and stellar mass range. Hence, these galaxies are not completely devoid
of star formation activity, and in Sect.~\ref{sec:gmf} we showed that the expected gas mass fraction, taking their lower sSFR values
into account, is higher than the upper limits obtained from the stacked images. It thus appears that the  effect
of a relatively modest reduction in the specific star formation rate has a large effect on the dust emission. It is not clear whether these
galaxies are transitioning to a truly quiescent stage, or if they are in a low-activity phase of episodic star formation activity.
The high average S\'{e}rsic index of the non-detected $z\sim2$ and $z\sim3$ galaxies (Figure~\ref{fig:scatter_1}) could suggest that they
are transitioning to become a part of the truly quiescent population.

\medskip

None of our galaxies in the $z\sim4$ and $z\sim5$ samples are detected with ALMA and the upper limits to the gas mass fractions derived from
stacked images (0.38 and 0.37, respectively) are much lower than what is expected from scaling relations. They are, however, not dead galaxies
in terms of their star formation activity, as seen through their SFRs. Their average star formation rate per unit stellar mass shows that they are
as efficient in forming stars as their lower redshift counterparts (Figure~\ref{fig:ssfr} and~\ref{fig:scatter_2}). The difference is that the $z\gtrsim4$ galaxies have
stellar masses $\sim$10 times lower than the $z$=2 sample. Defining the star formation efficiency (SFE) as $SFR/M_{\mathrm{gas}}$ in units
of Gyr$^{-1}$, we find that the submm detected galaxies have $SFE \sim 2-3$, while the $z$=4 and 5 samples have $SFE > 4$. The undetected
$z$=2 and 3 galaxies have $SFE > 1$. Since stars are formed from interstellar gas, this suggests that either the star formation process is 2-4 times
more efficient at $z\gtrsim4$, using up the interstellar gas faster than it can be replenished, or the total amount of interstellar gas in $z\gtrsim4$
galaxies is not reflected in the dust continuum, at least not in the stellar mass ranges probed in this study. The inverse of the SFE is the gas depletion
time scale, corresponding to $t_{\mathrm{depl}}\sim0.3-0.5$ Gyr and $<0.25$ Gyr for the submm detected galaxies and $z\gtrsim4$ galaxies,
respectively.

\medskip

As discussed in Sect.~\ref{sec:results}, there is an intrinsic bias against finding dust-obscured galaxies at $z\gtrsim4$ using optical
and near-infrared selected catalogs. This certainly applies to the CANDELS GOODS-S catalog which is based on H-band selected
sources. At $z\gtrsim4$ the H-band corresponds to rest-frame UV wavelengths and the presence of dust obscuration in low-mass galaxies
at these redshifts, with corresponding low luminosities, may simply make them too faint to be included in the catalog. Such galaxies should
show up in blind submm surveys as submm emission with no obvious optical/NIR counterpart. However, the fact that only a single $z\gtrsim3$ galaxy was found
in the 4.5 armin$^2$ ALMA continuum survey of the Hubble Ultra Deep Field (HUDF; Dunlop et al. 2017), none in the very deep 1.6
arcmin$^2$ survey of the HUDF (Aravena et al. 2016), and none in an ALMA survey
of three of the Frontier Fields (Laporte et al. 2017), argues against a `hidden population' of dusty, far-infrared luminous, star forming low-mass
galaxies at high redshift. The single exception in the UDF is CANDELS ID 12781 (Dunlop et al. 2017), which is a faint galaxy with a photometric
redshift $z_{\mathrm{phot}}=4.8$ and $M_{*}=3.1\times10^{9}$\,M$_{\odot}$.

\subsection{Metallicity and stellar mass}\label{sec:metallicity}

The gas-phase metallicity correlates with the stellar mass of a galaxy, with more massive galaxies having higher
metallicities than lower mass galaxies (e.g. Tremonti et al. 2004). This mass-metallicity relation (MZR) exists up to at least $z\sim3.5$
(e.g. Savaglio et al. 2005; Erb et al. 2006; Maiolino et al. 2008; Grasshorn Gebhardt et al., 2016; Guo et al. 2016; Sanders et al. 2018).
The conversion factor $\alpha_{850\mu m}$ used in this paper to derive the gas masses is empirically derived for galaxies with $\log{M_*/M_{\odot}}
> 10.2$ (Scoville et al. 2016). Our $z$=2--3 galaxies are all more massive, with an average stellar mass $\log{M_*/M_{\odot}}\sim10.6$.
Hence, their gas-phase metallicity should be comparable to those galaxies used to derive $\alpha_{850\mu m}$. The average stellar masses
of the $z$=4 and $z$=5 galaxies, however, are $\log{M_*/M_{\odot}} = 9.8$ and $9.7$. Assuming that the MZR can be extended to these
high redshifts, the gas-phase metallicity of our $z\gtrsim4$ galaxies can be as low as $\sim0.2$\,$Z_{\odot}$ (e.g. Genzel et al. 2015;
Tacconi et al. 2018), possibly affecting the gas-to-dust mass ratio and, in extension, the conversion factor $\alpha_{850\mu m}$.

\medskip

We can derive a correction to the $\alpha_{850\mu m}$ conversion factor by using the relation between gas-to-dust mass ratio
and gas-phase metallicity derived by Draine et al. (2007) and R\'{e}my-Ruyer et al. (2016).
Draine et al. (2007) derived gas-to-dust mass ratios for a sample of local galaxies with metallicities ranging from 0.08-1.0
$Z_{\odot}$\footnote{Assuming a solar metallicity $12 + \log{(O/H)} = 8.69$ (Asplund et al. 2009)} and found a linear relation
between gas phase metallicity and gas-to-dust mass ratio. The change in gas-to-dust mass ratio going from $Z/Z_{\odot}=1.0$ to 0.25
is $\sim$2.5, with a relatively small dispersion (see Fig. 7 in Draine et al. 2007). Below 0.25\,$Z_{\odot}$ ($12 + \log{(O/H)}
= 8.1$) the dispersion in the gas-to-dust mass ratio increases, with some galaxies having a lower than expected ratio. R\'{e}my-Ruyer
et al. (2016) presented an extensive study of the gas-to-dust ratios of local galaxies with metallicities ranging from $Z/Z_{\odot} = 0.03-2.0$
and found that for $Z>0.20$\,Z$_{\odot}$, the gas-to-dust ratio follows a linear correlation with gas-phase metallicity, similar to Draine
et al. (2007). Below $Z=0.20$\,Z$_{\odot}$, the gas-to-dust ratio has a steeper, but still linear, correlation with  the gas-phase metallicity. 
Using the linear relation between gas-phase metallicity and gas-to-dust mass ratio derived by Draine et al. (2007) and R\'{e}my-Ruyer et al.
(2014), we conclude that the gas masses of the $z\gtrsim4$ galaxies in our sample could be underestimated by a factor $\sim$3, compared to the more metal-rich
systems used to derive the relation between $L_{850\mu m}$ and gas mass (Scoville et al. 2016).
If this is the case, the upper limit to the gas mass fraction of the $z$=4 and $z$=5 samples would increase to $\sim$0.66. This value is similar
to the derived $f_{\mathrm{gas}}$ for the $z$=3 sample but far less than the $f_{\mathrm{gas}} \approx 0.9$ implied by the scaling relations.

\medskip

It is interesting to compare the $z$=4.4 sample of Scoville et al. (2016) with our $z$=4 and 5 samples. Scoville et al. (2016)report a 60\%
detection rate of dust continuum emission, while we  have a 0\% detection rate for our $z\gtrsim4$ galaxies. The observing strategy and
sensitivity limits are comparable. The average stellar mass of the Scoville et al. galaxies with dust emission is $\log{M_{*}/M_{\odot}} = 10.89$,
with an average gas mass fraction of $f_{\mathrm{gas}}\sim0.68$. This value includes the effect of the CMB (see Sect.~\ref{sec:submm}).
So why do Scoville et al. (2016) have a detection rate of 60\% while the detection rate in our sample is 0\%? The only difference between
the samples is the stellar mass, with our galaxies on average being 15 times less massive than those of Scoville et al. (2016).

The gas mass fraction is supposed to be higher for less massive galaxies compared to higher mass galaxies, but a lower gas-phase
metallicities for the low-mass galaxies may make this gas more difficult to detect using dust emission as well as CO emission.
The average gas mass fraction for the submm detected galaxies in the Scoville et al. (2016) $z$=4.4 sample is consistent with expectations
from scaling laws for that stellar mass range (Figure~\ref{fig:scoville}). The upper limit to the gas mass fraction for our $z\gtrsim4$ galaxies,
corrected for a lower gas-phase metallicity, is $f_{\mathrm{gas}} = 0.66$, but for this stellar mass range the scaling laws predict a gas mass
fraction $\sim$0.85--0.90.

\medskip

The MZR has been shown to also depend on the star formation rate, or more specifically, sSFR (e.g. Ellison et al. 2008). The M$_*$-SFR-Z
relation shows that at fixed stellar mass, galaxies with higher SFR have lower gas-phase metallicities. The M$_*$-SFR-Z relation exists to at
least to $z\sim2.5$ (Mannucci et al. 2010; Sanders et al. 2018). The theoretical interpretation of this extended relation is that the star formation
is driven by the accretion of low-metallicity gas. When star formation is quenched, the dilution of low-metallicity gas ceases and the gas-phase
metallicity increases due to stellar mass loss.
Since SFR scales linearly with stellar mass, the M$_*$-SFR-Z relation would have a larger impact on the gas-phase metallicity for more massive
galaxies compared with lower-mass systems. Hence, metallicity alone does not seem to account for the non-detection of dust emission in our
$z\gtrsim4$ galaxies.

\medskip

{We have no observational input regarding the gas-phase metallicities of the galaxies in our sample and have to rely on extending the
MZR to $z\gtrsim4$.
The metallicities of our galaxies, listed in Table~\ref{tab:candelsdata} and shown in Figure~\ref{fig:scatter_3}, are derived from SED
fits using CANDELS photometry and only provide estimates of stellar metallicities. The results suggest that the average stellar
metallicity of the $z\gtrsim4$ galaxies is quite high, with $Z\sim0.9 \pm 0.5$\,$Z_{\odot}$, significantly higher than the estimates for
the gas-phase metallicity provided by the MZR. The metallicity derived from
SED fitting is subject to degeneracies with extinction and age. This can introduce an artificial scatter in the metallicity and extinction values. However,
no systematic effect has been found in the analysis of the CANDELS data made by several different teams using different assumptions about star
formation histories, extinction laws and different treatments of nebular emission, as well as the handling of photometric and systematic uncertainties
in the fitting process (e.g. Mobasher et al. 2015; Santini et al. 2015). The consistency of the high stellar metallicities in our sample thus indicates that
these estimates are quite robust when applied to the sample as a whole.

\subsection{Dust emission at high redshift}

The non-detection of dust emission from our $z\gtrsim4$ galaxies stands in contrast to a growing number of detections of dust emission as 
well as atomic fine-structure lines in seemingly normal $z\gtrsim 6$ galaxies. Tamura et al. (2018) detect both the [OIII]\,88$\mu$m line and
dust continuum in a gravitationally lensed galaxy at $z$=8.31. The de-lensed stellar mass is $M_{*} = 5 \times 10^{9}$\,M$_{\odot}$ and the
inferred dust mass is $4\times10^{6}$\,M$_{\odot}$. Both the dust and the [OIII] line emission coincide with the rest-frame UV emission on
a kpc scale. Walter et al. (2018) detect [OIII]\,88$\mu$m emission and dust continuum from a $z=$6.08 QSO as well as a neighboring galaxy,
allowing a direct comparison of the dust properties in a QSO host galaxy and a star forming galaxy at the same redshift. Dust continuum
emission was detected in a gravitationally lensed galaxy at $z$=7.5 (Watson et al. 2015; Knudsen et al. 2017). The de-lensed stellar mass
is $M_{*}=2\times10^{9}$\,M$_{\odot}$ and the estimated dust mass is $\sim1\times10^{7}$\,M$_{\odot}$. In all of these cases, the dust
emission is directly associated with the stellar component.  However, Maiolino et al. (2015) detected [CII]\,158$\mu$m emission from a
Lyman Break Galaxy (LBG) at $z$=7.10, but no dust emission. In this case, the [CII] emission is offset by $\sim$4\,kpc from the rest-frame
UV emission, possibly indicating that the dense gas in is rapidly being disrupted by stellar feedback processes.

\medskip

How do our $z\gtrsim4$ galaxies fit in with these $z\gtrsim6$ galaxies with large inferred dust masses? The $z\gtrsim6$
galaxies all have stellar masses of a few $10^{9}$\,M$_{\odot}$, enriched ISM with surprisingly large dust masses, and they
are forming stars at a high rate. Taken at face value, the high stellar metallicities of our $z\gtrsim4$ galaxies are consistent
with the rapid build-up of the stellar population seen in the $z\gtrsim6$ galaxies. In terms of cosmic time, $\sim$250--600\,Myr
separates our $z$=5 and $z$=4 galaxies from the $z\sim6$ galaxies. If the inferred SFRs are maintained over these times scales,
it is possible to build up a stellar mass of several $10^{10}$\,M$_{\odot}$. Hence, there might be a connection between the $z\gtrsim6$
dusty galaxies and the $z\sim4$ galaxies observed by Scoville et al. (2016). 
It is, however, also possible that the metal-enriched ISM is removed through stellar feedback. Galactic winds remove metals from
galaxies and do so with higher efficiency in low-mass galaxies due to their shallow potential wells (Dekel \& Silk 1986). Gas inflows
bring metal-poor gas into the galactic halo, diluting the metal content in the existing ISM (Ker\v{e}s et al. 2005; Faucher-Gigu\`{e}re
et al. 2011). This process could also leads to  re-accretion of some of the metal-enriched gas previously ejected via outflows (Bertone
et al. 2007; Oppenheimer et al. 2010). These sometimes competing processes could lead to a situation where the gas-phase metallicity
is lower than the stellar metallicities and where the gas-to-dust mass ratio is lower than expected for the stellar mass.
While only speculative, such a scenario would explain the non-detection of dust emission in the $z\gtrsim4$ galaxies in our sample.

\subsection{Detectability of dust at high redshift}

With these disparate observational results, a high detection rate of dust emission in $z\sim4$ galaxies with $\log{M_{\odot}/M_{*}}\sim10.8$,
no dust emission in $z\sim4$ galaxies with $\log{M_{\odot}/M_{*}}\sim9.8$ and the presence of galaxies at $z\gtrsim6$ with large dust
masses and vigorous star formation activity, we must ask where the dust comes from. Does it have a different composition compared to
lower redshift galaxies and can this affect the detectability of the dust emission in very high redshift galaxies?

There are three main sources of interstellar dust: condensation in SNe ejecta, producing mainly silicate type grains, ejecta
from AGB stars, producing mainly carbon type grains, and grain accretion processes in the dense ISM. Dust grains can be
destroyed via thermal sputtering, collisions with other dust grains and in SN shocks.
Grain growth in the ISM represents the dominant mode of dust formation in our Milky Way galaxy (e.g. Dwek 1998). Popping
et al. (2017) presented a study of the dust content of
galaxies up to $z$=9 using semi-analytic models (SAMs), including dust production, destruction and growth in the ISM. They
find that the ISM accretion mode is the dominant production channel for dust even for very high redshift galaxies, exceeding
the production rates from AGB stars and SNe by several orders of magnitude. Similar results have been reached by Dwek
et al. (2007) and Michalowski (2015) in analysis of the dust content of $z\gtrsim6$ galaxies and QSO hosts.

The stellar mass of our $z\gtrsim4$ galaxies is lower than the typical galaxies that have previously been part of dust continuum
studies at high redshift. The stellar mass is, however, high enough that we can reasonably expect a gas-phase metallicity of
at least $\gtrsim0.2$\,Z$_{\odot}$ and a gas-to-dust ratio within a factor of $\sim$2-3 of more massive and, hence, more metal
rich galaxies. This puts our $z\gtrsim4$ galaxies well within the detectable range of dust continuum emission from the stacked
images.
The lack of detections at $z\gtrsim4$ could potentially be due to different dust properties, leading to different emissivity and
observed dust continuum flux density. However, both theoretical modeling (e.g. Popping et al. 2017) and the fact that dust
is observed in more massive $z\gtrsim4$ galaxies and QSO hosts argues against such a scenario. As long as the evolution
time-scales of more massive galaxies are similar to the galaxies in our sample, the dust production channels ought to be similar,
leading to similar dust properties. Furthermore, since the stellar metallicities of the $z\gtrsim4$ galaxies in our sample appear to
be $\sim$0.9\,Z$_{\odot}$, their star formation histories cannot be too different from their more massive counterparts
showing similarly high stellar metallicities..

Finally, in our analysis we have assumed a dust temperature of 25\,K. The inferred dust temperature from submm
observations of $z\gtrsim4$ SMGs and QSO host galaxies is usually higher. The environments in these objects
could be different from that of lower mass galaxies at the same redshift, providing a higher flux of UV photons
and heating the dust to a higher temperature. If we would assume a higher dust temperature in our analysis,
the inferred dust and gas masses, including the upper limits, would have to be adjusted downward. Clearly this 
does not help in explaining the non-detection of the $z\gtrsim4$ galaxies in our sample.

Hence, while it is fair to say that a significant uncertainty remains on the dust properties for high redshift galaxies,
the non-detections of dust emission from our $z\gtrsim4$ galaxies, and the corresponding upper limits to the gas
mass fraction, remains a conundrum. It seems more likely to be related to dynamical effects in the ISM, such as
removal of the interstellar gas after an initial star formation phase, and subsequent dilution of the ISM by metal-poor
gas, rather than an intrinsic low gas-phase metallicity or a high dust destruction rate.

\section{Summary and conclusions}\label{sec:summary}

We present an ALMA survey of dust continuum emission in a sample of 70 galaxies in the redshift range
$z$=2-5. The sample contains galaxies connected through an evolutionary sequence representing how
galaxies grow in mass with cosmic time. Multi-Epoch Abundance Matching (MEAM) is used to define the
sample of likely progenitors of a $z$=0 galaxy of stellar mass $1.5 \times 10^{11}$\,M$_{\odot}$, seen at
different epochs. The selection takes the stochastic nature of galaxy growth into account. {\it No other criteria
apart from redshift and stellar mass were used in selecting the galaxies our sample}.
We obtained ALMA band 7 ($z$$\sim$2 and 3) and band 6 ($z$$\sim$4 and 5) observations of the dust
continuum and used an empirically derived conversion factor between specific 850$\mu$m luminosity
and gas mass to convert the observed fluxes into an estimate of the gas mass. Ancillary data from the
CANDELS GOODS-S survey are used to derive the gas mass fractions and correlate this with other
parameters of the target galaxies. We define the gas mass fraction as
$f_{\mathrm{gas}}  = M_{\mathrm{gas}}/(M_{\mathrm{gas}} + M_{*})$. All gas masses and gas mass
fractions have been corrected for an increasing temperature of the Cosmic Microwave Background.
Upper limits are quoted at 3$\sigma$.

The main results from this study are:

\begin{enumerate}

\item The detection rate for the $z$=2 redshift bin is 75\%, while it is 50\% for the $z$=3 bin and 0\% for the $z$=4 and 5 redshift
bins.

\item The average gas mass fraction for the $z$=2 redshift bin is $f_{\mathrm{gas}} = 0.55 \pm 0.12$ and for the $z$=3 bin
$f_{\mathrm{gas}} = 0.62 \pm 0.15$, only taking the detected galaxies into account.

\item Stacked images of the galaxies in our sample not detected with ALMA in the $z$=2 and $z$=3 redshift bins provide 3$\sigma$ upper limits
to $f_{\mathrm{gas}} < 0.08$ and 0.15, respectively.
Stacked images of the $z$=4 and $z$=5 galaxies reach a rms noise of $\sim$8\,$\mu$Jy and the 3$\sigma$ upper limits
correspond to $f_{\mathrm{gas}} < 0.38$ and 0.37, respectively. Correcting for a lower gas-phase metallicity
increases these upper limits to $\lesssim$0.66.

\item Comparison with several different scaling relations show a good correspondence between the observed average gas
mass fraction for our $z\lesssim3$ galaxies with detectable submm emission.

\item For $z\lesssim3$, the observed gas mass fraction decreases linearly with cosmic time at a rate $0.043\pm0.007$\,Gyr$^{-1}$.

\item The gas-phase metallicity can effect the estimated gas mass fractions for the highest redshift bins, but even taking this into account,
metallicity alone cannot explain the low upper limits for the $z$=4 and $z$=5 samples.

\item The metallicities of the stellar population is high, with $\sim$0.9\,$Z_{\odot}$ for the $z\gtrsim4$ galaxies, possibly suggesting
a disconnect between the current gas-phase and stellar metallicities for the highest redshift galaxies.

\end{enumerate}

Combining our results with gas mass estimates from the literature for galaxies at redshifts $z\lesssim3$ shows that
the gas mass fraction increases dramatically from $z$=0 to $z$=3. The gas mass fraction at $z\sim3$ is $\sim$8 times
higher than at $z\sim0$ for galaxies of comparable stellar mass.
The non-detected galaxies in our sample appear to have very low levels of interstellar gas.
At $z\gtrsim4$ the gas-phase metallicity
could be low enough to affect the gas-to-dust mass ratio by a factor $\sim$3. Should this be the case, the observed upper limits to
the gas mass fractions are higher but still too low to be compatible with what is expected from scaling relations. The stellar metallicity,
derived from SED fits, appear to be high for the $z=4-5$ galaxies, with an average metallicity of 0.9\,$Z_{\odot}$. This is
much higher than the expected gas-phase metallicity of $\sim0.2$\,$Z_{\odot}$. One scenario that can explain this large
discrepancy is that the $z\gtrsim4$ galaxies are undergoing a second accretion phase of primordial or low-metallicity
gas, and that the existing stellar population was build up at an earlier stage where the ISM was either used up or expelled
from the galaxy.

The lack of detection of dust continuum emission from the $z\gtrsim4$ galaxies is in contrast with results from 
a survey targeting significantly more massive galaxies at $z\sim4.4$.
If this is due to a stellar mass dependence, the correlation between stellar mass and gas mass is different from
that of galaxies at lower redshifts, where the gas mass fraction increases with lower stellar mass.
A longitudinal study of the gas content of galaxies at $z\sim4$ would be valuable to determine at which
stellar mass the dust emission becomes present. A direct comparison of the gas-phase and stellar metallicities
of galaxies at $z\gtrsim4$ will be possible once JWST is launched.

\acknowledgments
This paper makes use of the following ALMA data: ADS/JAO.ALMA\#2015.1.00870.S. ALMA is a partnership of ESO (representing its member states),
NSF (USA) and NINS (Japan), together with NRC (Canada), MOST and ASIAA (Taiwan), and KASI (Republic of Korea), in cooperation with the Republic
of Chile. The Joint ALMA Observatory is operated by ESO, AUI/NRAO and NAOJ. The National Radio Astronomy Observatory is a facility of the National
Science Foundation operated under cooperative agreement by Associated Universities, Inc. 
Support for the CANDELS Program HST-GO-12060 was provided by NASA through a grant from the Space Telescope Science Institute, which is operated
by the Association of Universities for Research in Astronomy, Incorporated, under NASA contract NAS5-26555.
G.B.B acknowledges support from The Cosmic Dawn Center, which is funded by the Danish National Research Foundation.

\startlongtable
\begin{deluxetable}{cccccccccc}
\tablewidth{0pt}
\tabletypesize{\scriptsize}

\tablecaption{ALMA data for the targets\label{tab:almadata}}

\tablehead{
\ \\
\multicolumn{1}{c}{(1)}      &
\multicolumn{1}{c}{(2)}      &
\multicolumn{1}{c}{(3)}      &
\multicolumn{1}{c}{(4)}      &
\multicolumn{1}{c}{(5)}      &
\multicolumn{1}{c}{(6)}      &
\multicolumn{1}{c}{(7)}      &
\multicolumn{1}{c}{(8)}      &
\multicolumn{1}{c}{(9)}      &
\multicolumn{1}{c}{(10)}     \\
\multicolumn{1}{c}{CANDELS ID}                                           &
\multicolumn{1}{c}{RA}                                                            &
\multicolumn{1}{c}{DEC}                                                         &
\multicolumn{1}{c}{z}                                                              &
\multicolumn{1}{c}{$S_{\nu}$}                                                 &
\multicolumn{1}{c}{$\sigma$}                                                  &
\multicolumn{1}{c}{SNR}                                                         &
\multicolumn{1}{c}{$\alpha_{\mathrm{CMB}}$}                       &
\multicolumn{1}{c}{$\log{M_{\mathrm{gas}}/M_{\odot}}$}       &
\multicolumn{1}{c}{$f_{\mathrm{gas}}$}                                  \\
\multicolumn{1}{c}{}                                            &
\multicolumn{2}{c}{J2000.0}                               &
\multicolumn{1}{c}{}                                            &
\multicolumn{1}{c}{mJy}                                     &
\multicolumn{1}{c}{$\mu$Jy/beam}                    &
\multicolumn{1}{c}{}                                            &
\multicolumn{1}{c}{}                                            &
\multicolumn{1}{c}{}                                            &
\multicolumn{1}{c}{}                                            \\
}
\startdata
{\bf z = 2} \\
Detections  \\
  2619 & 53.1635384 & -27.8904751 & 2.447$^{\ }$ & 2.409 & 90 &   26.73 & 1.02 & 11.162 &  0.61 \\
 2344 & 53.0742341 & -27.8932888 & 2.378$^{\ }$ & 0.417 & 94 &    4.46 & 1.02 & 10.400 &  0.35 \\
 3280 & 53.0606151 & -27.8823723 & 2.155$^{\ }$ & 1.124 & 93 &   12.07 & 1.02 & 10.832 &  0.53 \\
 7034 & 53.1020377 & -27.8461169 & 1.997$^{\ }$ & 0.588 & 79 &    7.42 & 1.01 & 10.553 &  0.33 \\
 7670 & 53.1481791 & -27.8391622 & 1.977$^{\ }$ & 1.128 & 95 &   11.92 & 1.01 & 10.836 &  0.59 \\
10973 & 53.1858293 & -27.8099656 & 2.583$^*$ & 0.851 & 89 &    9.57 & 1.02 & 10.710 &  0.57 \\
12363 & 53.1202141 & -27.7988536 & 2.427$^{\ }$ & 0.743 & 92 &    8.10 & 1.02 & 10.651 &  0.51 \\
12537 & 53.0568179 & -27.7982999 & 1.742$^*$ & 1.310 & 87 &   15.14 & 1.01 & 10.903 &  0.61 \\
15586 & 53.0536210 & -27.7780356 & 1.885$^{\ }$ & 0.740 & 87 &    8.49 & 1.01 & 10.654 &  0.39 \\
16972 & 53.0819572 & -27.7672022 & 1.618$^{\ }$ & 5.228 & 90 &   57.95 & 1.01 & 11.505 &  0.87 \\
19476 & 53.0835369 & -27.7464307 & 1.897$^*$ & 0.591 & 84 &    7.06 & 1.01 & 10.556 &  0.48 \\
20253 & 53.1929134 & -27.7383620 & 2.434$^*$ & 1.421 & 94 &   15.17 & 1.02 & 10.933 &  0.74 \\
22852 & 53.1753092 & -27.6947705 & 2.440$^{\ }$ & 5.058 & 90 &   55.90 & 1.02 & 11.484 &  0.76 \\
23090 & 53.0860566 & -27.7095666 & 1.861$^{\ }$ & 2.566 & 92 &   27.82 & 1.01 & 11.194 &  0.68 \\
25928 & 53.1736928 & -27.6980881 & 2.082$^{\ }$ & 0.546 & 80 &    6.87 & 1.02 & 10.520 &  0.31 \\
\ \\
Non-detections \\
 1119 & 53.1690525 & -27.9158458 & 2.059$^{\ }$ & --- & 85 & --- & 1.05 & $<$10.733 &  $<$0.35 \\
 2552 & 53.1240874 & -27.8912028 & 2.459$^{\ }$ & --- & 82 & --- & 1.07 & $<$10.676 &  $<$0.31 \\
 4337 & 53.1161464 & -27.8719167 & 1.874$^{\ }$ & --- & 81 & --- & 1.05 & $<$10.734 &  $<$0.50 \\
11368 & 53.1165089 & -27.8067456 & 2.283$^{\ }$ & --- & 87 & --- & 1.06 & $<$10.720 &  $<$0.42 \\
26496 & 53.0206631 & -27.7008296 & 2.097$^{\ }$ & --- & 87 & --- & 1.06 & $<$10.740 &  $<$0.58 \\
\ \\
\hline
\ \\
{\bf z = 3} \\
Detections \\
  372 & 53.0929203 & -27.9363229 & 2.695$^*$ & 1.104 & 70 &   15.87 & 1.03 & 10.823 &  0.61 \\
 2701 & 53.1463516 & -27.8887610 & 2.970$^{\ }$ & 2.669 & 91 &   29.27 & 1.03 & 11.210 &  0.69 \\
 4438 & 53.2151604 & -27.8702368 & 3.096$^*$ & 0.847 & 92 &    9.25 & 1.03 & 10.714 &  0.59 \\
 4878 & 53.1354720 & -27.8662145 & 3.150$^{\ }$ & 1.300 & 92 &   14.12 & 1.04 & 10.901 &  0.77 \\
 8433 & 53.0710315 & -27.8324895 & 3.334$^{\ }$ & 0.427 & 90 &    4.75 & 1.04 & 10.423 &  0.48 \\
 9286 & 53.2243550 & -27.8243357 & 3.243$^{\ }$ & 1.065 & 89 &   11.90 & 1.04 & 10.817 &  0.70 \\
10832 & 53.1614111 & -27.8110800 & 2.795$^{\ }$ & 1.030 & 97 &   10.59 & 1.03 & 10.794 &  0.65 \\
11659 & 53.1072811 & -27.8040705 & 2.718$^{\ }$ & 0.830 & 93 &    8.89 & 1.03 & 10.700 &  0.56 \\
14781 & 53.0333278 & -27.7825748 & 2.619$^*$ & 0.533 & 96 &   89.16 & 1.02 & 10.507 &  0.35 \\
19692 & 53.0552398 & -27.7433425 & 2.956$^{\ }$ & 1.347 & 92 &   14.72 & 1.03 & 10.913 &  0.71 \\
22281 & 53.0972767 & -27.7203525 & 3.046$^{\ }$ & 1.822 & 90 &   20.25 & 1.03 & 11.046 &  0.70 \\
\ \\
Non-detections \\
 1424 & 53.1937574 & -27.9082656 & 2.979$^{\ }$ & --- & 90 & --- & 1.09 & $<$10.662 &  $<$0.40 \\
 2782 & 53.0835712 & -27.8875292 & 3.472$^{\ }$ & --- & 86 & --- & 1.12 & $<$10.603 &  $<$0.50 \\
 2807 & 53.1515562 & -27.8870850 & 3.581$^{\ }$ & --- & 88 & --- & 1.13 & $<$10.607 &  $<$0.62 \\
 3448 & 53.2417433 & -27.8798285 & 2.669$^*$ & --- & 91 & --- & 1.08 & $<$10.698 &  $<$0.51 \\
 6780 & 53.0745639 & -27.8472601 & 3.495$^*$ & --- & 95 & --- & 1.13 & $<$10.645 &  $<$0.59 \\
 7526 & 53.0786782 & -27.8395462 & 3.422$^{\ }$ & --- & 94 & --- & 1.12 & $<$10.645 &  $<$0.68 \\
 8339 & 53.1701864 & -27.8333559 & 3.530$^{\ }$ & --- & 92 & --- & 1.13 & $<$10.632 &  $<$0.72 \\
19505 & 53.0166020 & -27.7448468 & 3.331$^{\ }$ & --- & 90 & --- & 1.11 & $<$10.637 &  $<$0.71 \\
22211 & 52.9988126 & -27.7209744 & 2.955$^{\ }$ & --- & 97 & --- & 1.09 & $<$10.699 &  $<$0.62 \\
\ \\
\hline
\ \\
{\bf z = 4} \\
Non-detections \\
 1479 & 53.2021741 & -27.9071077 & 4.391$^{\ }$ & --- & 37 & --- & 1.21 &  $<$9.864 &  $<$0.82 \\
 2663 & 53.1161215 & -27.8889058 & 4.264$^{\ }$ & --- & 34 & --- & 1.19 &  $<$9.815 &  $<$0.62 \\
 2997 & 53.1997842 & -27.8849372 & 4.319$^{\ }$ & --- & 32 & --- & 1.20 &  $<$9.795 &  $<$0.43 \\
 3753 & 53.0792917 & -27.8772595 & 4.431$^*$ & --- & 34 & --- & 1.21 &  $<$9.830 &  $<$0.37 \\
 3962 & 53.0678856 & -27.8745094 & 3.813$^*$ & --- & 35 & --- & 1.15 &  $<$9.813 &  $<$0.47 \\
 5817 & 53.0707573 & -27.8564182 & 4.643$^{\ }$ & --- & 36 & --- & 1.24 &  $<$9.867 &  $<$0.44 \\
 7188 & 53.1981318 & -27.8424852 & 4.488$^{\ }$ & --- & 36 & --- & 1.22 &  $<$9.850 &  $<$0.76 \\
 8178 & 53.0742628 & -27.8337447 & 4.530$^{\ }$ & --- & 36 & --- & 1.22 &  $<$9.860 &  $<$0.78 \\
11480 & 53.0385920 & -27.8051393 & 4.303$^{\ }$ & --- & 38 & --- & 1.20 &  $<$9.871 &  $<$0.52 \\
12025 & 53.0533002 & -27.8007117 & 4.394$^{\ }$ & --- & 36 & --- & 1.21 &  $<$9.846 &  $<$0.55 \\
13214 & 53.1856931 & -27.7920380 & 4.386$^{\ }$ & --- & 37 & --- & 1.21 &  $<$9.864 &  $<$0.38 \\
14308 & 53.0618828 & -27.7850707 & 4.442$^*$ & --- & 34 & --- & 1.21 &  $<$9.823 &  $<$0.65 \\
17454 & 53.2122742 & -27.7629709 & 4.336$^{\ }$ & --- & 30 & --- & 1.20 &  $<$9.772 &  $<$0.59 \\
18554 & 53.0748825 & -27.7534684 & 3.912$^{\ }$ & --- & 32 & --- & 1.16 &  $<$9.776 &  $<$0.31 \\
19842 & 53.2060773 & -27.7416201 & 3.979$^{\ }$ & --- & 33 & --- & 1.16 &  $<$9.794 &  $<$0.49 \\
21506 & 53.2043708 & -27.7267362 & 4.212$^{\ }$ & --- & 38 & --- & 1.19 &  $<$9.868 &  $<$0.49 \\
25875 & 53.1494101 & -27.6973764 & 4.379$^{\ }$ & --- & 35 & --- & 1.21 &  $<$9.833 &  $<$0.62 \\
26960 & 53.1749626 & -27.9342009 & 4.444$^{\ }$ & --- & 34 & --- & 1.21 &  $<$9.830 &  $<$0.63 \\
31211 & 53.0313864 & -27.7847046 & 4.530$^{\ }$ & --- & 39 & --- & 1.22 &  $<$9.892 &  $<$0.85 \\
33501 & 53.0434199 & -27.7236716 & 4.591$^{\ }$ & --- & 37 & --- & 1.23 &  $<$9.877 &  $<$0.71 \\
\ \\
\hline
\ \\
{\bf z = 5} \\
Non-detections \\
 6741 & 53.1307981 & -27.8468178 & 5.477$^{\ }$ & --- & 25 & --- & 1.38 &  $<$9.760 &  $<$0.46 \\
 7310 & 53.1394807 & -27.8416664 & 4.948$^*$ & --- & 29 & --- & 1.28 &  $<$9.786 &  $<$0.33 \\
10595 & 53.0403963 & -27.8131821 & 4.961$^{\ }$ & --- & 25 & --- & 1.28 &  $<$9.720 &  $<$0.26 \\
12383 & 53.0731951 & -27.7978582 & 4.796$^{\ }$ & --- & 26 & --- & 1.26 &  $<$9.734 &  $<$0.75 \\
17427 & 53.1036877 & -27.7631692 & 4.893$^{\ }$ & --- & 25 & --- & 1.27 &  $<$9.720 &  $<$0.71 \\
19247 & 53.0324708 & -27.7472779 & 4.974$^{\ }$ & --- & 24 & --- & 1.29 &  $<$9.704 &  $<$0.75 \\
21665 & 53.0399591 & -27.7255058 & 4.960$^{\ }$ & --- & 24 & --- & 1.28 &  $<$9.716 &  $<$0.74 \\
23926 & 53.1596602 & -27.6611462 & 5.129$^{\ }$ & --- & 25 & --- & 1.31 &  $<$9.738 &  $<$0.44 \\
32882 & 53.0153517 & -27.7426846 & 4.659$^{\ }$ & --- & 25 & --- & 1.24 &  $<$9.705 &  $<$0.58 \\
33659 & 53.1523454 & -27.7194612 & 4.991$^{\ }$ & --- & 25 & --- & 1.29 &  $<$9.728 &  $<$0.82 \\
\enddata
\tablecomments{Spectroscopic redshifts are marked with an asterisk. All other are photometric redshifts from the CANDELS catalog (Santini et al. 2015).}
\tablecomments{Coordinates are from the CANDELS H-band selected catalog.}
\tablecomments{The values listed for $f_{\mathrm{gas}}$ and $M_{\mathrm{gas}}$ have been corrected for the effects
of a higher Cosmic Microwave Background temperature (da Cunha et al. (2013) with the factor listed as $\alpha_{\mathrm{CMB}}$.}
\tablecomments{The $M_{\mathrm{gas}}$ and $f_{\mathrm{gas}}$ values for non-detections represent 3$\sigma$ upper limits.}
\end{deluxetable}

\newpage
\startlongtable
\begin{deluxetable*}{ccccccccccc}
\tablewidth{0pt}
\tabletypesize{\scriptsize}

\tablecaption{CANDELS data for the targets\label{tab:candelsdata}}

\tablehead{
\\
\multicolumn{1}{c}{(1)}      &
\multicolumn{1}{c}{(2)}      &
\multicolumn{1}{c}{(3)}      &
\multicolumn{1}{c}{(4)}      &
\multicolumn{1}{c}{(5)}      &
\multicolumn{1}{c}{(6)}      &
\multicolumn{1}{c}{(7)}      &
\multicolumn{1}{c}{(8)}      &
\multicolumn{1}{c}{(9)}      &
\multicolumn{1}{c}{(10)}    &
\multicolumn{1}{c}{(11)}    \\
\ \\
\multicolumn{1}{c}{CANDELS ID}                       &
\multicolumn{1}{c}{z}                                          &
\multicolumn{1}{c}{AGN Flag\tablenotemark{(1)}}                            &
\multicolumn{1}{c}{$\log{M_{*}/M_{\odot}}$}       &
\multicolumn{1}{c}{SFR\tablenotemark{(2)}}       &
\multicolumn{1}{c}{sSFR\tablenotemark{(2)}}     &
\multicolumn{1}{c}{E$_{\mathrm{B-V}}$}            &
\multicolumn{1}{c}{$\log{t_{\mathrm{age}}}$}     &
\multicolumn{1}{c}{Z}                                          &
\multicolumn{1}{c}{Sersic}                                  &
\multicolumn{1}{c}{R$_{e}$}                               \\
\multicolumn{3}{c}{}                                          &
\multicolumn{1}{c}{}                                          &
\multicolumn{1}{c}{M$_{\odot}$\,yr$^{-1}$}      &
\multicolumn{1}{c}{Gyr$^{-1}$}                         &
\multicolumn{1}{c}{ }                                         &
\multicolumn{1}{c}{yrs}                                     &
\multicolumn{1}{c}{Z$_{\odot}$}                       &
\multicolumn{1}{c}{ }                                         &
\multicolumn{1}{c}{kpc}                                \\
}
\startdata
{\bf z = 2} \\
Detections \\
  2619 & 2.447 & 0 &10.974 & 140.13 &  1.49 & 0.350 & 9.158 & 1.77 &  2.03 &  3.56 \\
 2344 & 2.378 & 0 &10.678 &  67.51 &  1.41 & 0.590 & 9.198 & 0.64 &  0.75 &  1.88 \\
 3280 & 2.155 & 0 &10.787 &  88.73 &  1.46 & 0.350 & 9.122 & 1.22 &  3.75 &  0.76 \\
 7034 & 1.997 & 0 &10.862 & 107.51 &  1.47 & 0.480 & 8.981 & 1.11 &  0.62 &  5.52 \\
 7670 & 1.977 & 0 &10.677 & 222.32 &  4.63 & 0.800 & 8.761 & 1.44 &  1.00 &  3.60 \\
10973 & 2.583 & 1 &10.582 &  52.33 &  1.38 & 0.220 & 9.245 & 2.10 &  1.94 &  0.49 \\
12363 & 2.427 & 0 &10.638 &  68.56 &  1.59 & 0.400 & 8.794 & 1.56 &  0.57 &  3.83 \\
12537 & 1.742 & 0 &10.712 &  96.55 &  1.89 & 0.430 & 8.921 & 1.01 &  1.45 &  5.90 \\
15586 & 1.885 & 0 &10.853 &  48.72 &  0.69 & 0.680 & 9.022 & 1.83 &  1.17 &  5.76 \\
16972 & 1.618 & 1 &10.665 & 143.94 &  3.13 & 0.870 & 8.947 & 1.89 &  0.95 &  7.56 \\
19476 & 1.897 & 1 &10.598 &  90.84 &  2.27 & 0.360 & 8.715 & 1.37 &  8.00 &  3.72 \\
20253 & 2.434 & 0 &10.486 & 106.80 &  3.44 & 0.340 & 8.858 & 0.99 &  1.44 &  2.98 \\
22852 & 2.440 & 0 &10.995 & 174.15 &  1.76 & 0.840 & 8.796 & 1.99 &  1.09 &  3.24 \\
23090 & 1.861 & 0 &10.867 & 218.79 &  2.96 & 0.580 & 8.785 & 0.66 &  1.16 &  5.58 \\
25928 & 2.082 & 0 &10.864 & 182.24 &  2.50 & 0.380 & 8.742 & 0.92 &  4.84 & 10.40 \\
\ \\
Non-detections \\
 1119 & 2.059 & 0 &11.004 &  47.01 &  0.47 & 0.240 & 9.090 & 0.87 &  3.88 &  1.29 \\
 2552 & 2.459 & 1 &11.021 &  30.22 &  0.28 & 0.440 & 8.954 & 1.05 &  3.29 &  4.37 \\
 4337 & 1.874 & 0 &10.741 &  39.74 &  0.72 & 0.220 & 9.219 & 1.15 &  2.69 &  1.48 \\
11368 & 2.283 & 0 &10.859 &  48.69 &  0.68 & 0.180 & 8.983 & 0.86 &  2.85 &  0.73 \\
26496 & 2.097 & 0 &10.596 & 260.63 &  6.68 & 0.390 & 8.429 & 0.76 &  0.61 &  3.78 \\
\ \\
\hline
Averages & z = 2 \\
\hline
All                    & 2.135 & & 10.800 & 118 & 2.38 & 0.46 & 8.98 & 1.26 & 1.9 & 3.82 \\
Detections       & 2.128 & & 10.773 & 107 & 1.76 & 0.51 & 8.97 & 1.37 & 1.6 & 4.32 \\
Non-detection & 2.154 & & 10.872 &   47 & 0.68 & 0.29 & 9.00 & 0.94 & 2.7 & 2.33 \\
\hline
\ \\
{\bf z = 3} \\
Detections \\
  372 & 2.695 & 0 &10.628 & 120.43 &  2.80 & 0.260 & 8.899 & 1.65 &  1.20 &  4.84 \\
 2701 & 2.970 & 0 &10.863 & 254.20 &  3.48 & 0.780 & 8.621 & 1.62 &  0.20 &  3.68 \\
 4438 & 3.096 & 1 &10.556 & 923.49 & 25.65 & 0.730 & 8.527 & 1.44 &  1.45 &  1.94 \\
 4878 & 3.150 & 0 &10.382 & 509.44 & 21.23 & 0.510 & 8.255 & 0.90 &  2.32 &  0.89 \\
 8433 & 3.334 & 0 &10.464 & 115.46 &  3.98 & 0.240 & 8.737 & 1.15 &  0.50 &  2.21 \\
 9286 & 3.243 & 0 &10.442 &  34.22 &  1.22 & 0.440 & 9.068 & 0.73 &  1.15 &  2.43 \\
10832 & 2.795 & 0 &10.521 & 180.56 &  5.47 & 0.650 & 8.611 & 1.27 &  8.00 &  1.60 \\
11659 & 2.718 & 0 &10.593 &  93.47 &  2.40 & 0.570 & 8.760 & 1.27 &  0.53 &  3.73 \\
14781 & 2.619 & 1 &10.773 &  62.77 &  1.06 & 0.190 & 9.292 & 1.95 &  1.78 &  1.25 \\
19692 & 2.956 & 0 &10.532 &  23.10 &  0.68 & 0.270 & 9.218 & 1.95 &  0.98 &  1.62 \\
22281 & 3.046 & 0 &10.675 &  92.54 &  1.97 & 0.510 & 8.816 & 1.55 &  0.50 &  4.13 \\
\ \\
Non-detection \\
 1424 & 2.979 & 0 &10.831 &  19.05 &  0.28 & 0.260 & 9.063 & 0.92 &  1.44 &  0.78 \\
 2782 & 3.472 & 0 &10.608 &  18.28 &  0.46 & 0.180 & 8.991 & 0.24 &  8.00 &  0.66 \\
 2807 & 3.581 & 0 &10.391 &  18.42 &  0.74 & 0.220 & 9.048 & 0.40 &  6.17 &  1.68 \\
 3448 & 2.669 & 0 &10.685 &  12.02 &  0.25 & 0.190 & 9.158 & 0.62 &  1.73 &  0.76 \\
 6780 & 3.495 & 0 &10.491 &  96.57 &  3.12 & 0.080 & 8.501 & 0.52 &  1.46 &  1.63 \\
 7526 & 3.422 & 0 &10.316 &   6.62 &  0.31 & 0.190 & 9.053 & 0.34 &  5.18 &  0.85 \\
 8339 & 3.530 & 0 &10.225 & 103.66 &  6.10 & 0.090 & 8.369 & 0.54 &  1.70 &  2.40 \\
19505 & 3.331 & 1 &10.258 & 171.07 &  9.50 & 0.210 & 8.713 & 0.79 &  6.47 &  0.41 \\
22211 & 2.955 & 0 &10.491 &   8.48 &  0.27 & 0.230 & 9.088 & 0.50 &  1.68 &  0.40 \\
\ \\
\hline
Averages & z = 3 \\
\hline
All                    & 3.103 & & 10.571 &   43 & 0.73 & 0.34 & 8.92 & 1.02 & 2.0 & 1.89 \\
Detections       & 2.966 & & 10.607 & 115 & 2.80 & 0.47 & 8.90 & 1.41 & 1.1 & 2.57 \\
Non-detection & 3.270 & & 10.521 &   18 & 0.46 & 0.18 & 8.96 & 0.54 & 3.3 & 1.06 \\
\hline
\ \\
{\bf z = 4} \\
Non-detections \\
 1479 & 4.391 & 0 & 9.220 &  11.99 &  7.05 & 0.100 & 8.619 & 1.65 &  7.93 & 21.89 \\
 2663 & 4.264 & 0 & 9.602 &   9.76 &  2.44 & 0.070 & 8.672 & 0.47 &  0.52 &  1.54 \\
 2997 & 4.319 & 0 & 9.915 &  21.19 &  2.58 & 0.090 & 8.788 & 0.73 &  6.92 &  0.71 \\
 3753 & 4.431 & 0 &10.068 &  49.48 &  4.12 & 0.090 & 8.584 & 0.60 &  0.73 &  2.49 \\
 3962 & 3.813 & 0 & 9.860 &  43.68 &  6.07 & 0.330 & 8.414 & 0.66 &  1.11 &  0.89 \\
 5817 & 4.643 & 0 & 9.972 &  26.94 &  2.87 & 0.050 & 8.771 & 1.05 &  3.87 &  3.81 \\
 7188 & 4.488 & 0 & 9.346 &   5.72 &  2.60 & 0.020 & 8.632 & 0.26 &  8.00 &  4.15 \\
 8178 & 4.530 & 0 & 9.318 &   6.55 &  3.12 & 0.060 & 8.711 & 1.08 &  1.86 &  0.73 \\
11480 & 4.303 & 0 & 9.832 &  16.61 &  2.44 & 0.120 & 8.650 & 0.29 &  0.61 &  1.59 \\
12025 & 4.394 & 0 & 9.757 &  12.51 &  2.20 & 0.070 & 8.662 & 0.42 &  2.16 &  2.47 \\
13214 & 4.386 & 0 &10.068 &   6.01 &  0.50 & 0.120 & 9.031 & 1.79 &  0.95 &  0.84 \\
14308 & 4.442 & 0 & 9.553 &  27.41 &  7.61 & 0.070 & 8.447 & 1.12 &  8.00 &  2.13 \\
17454 & 4.336 & 0 & 9.612 &  49.91 & 12.17 & 0.150 & 8.395 & 1.80 &  3.30 &  0.44 \\
18554 & 3.912 & 0 &10.121 & 122.51 &  9.42 & 0.330 & 8.388 & 0.37 &  8.00 &  4.05 \\
19842 & 3.979 & 0 & 9.813 &  10.37 &  1.60 & 0.120 & 8.898 & 0.41 &  0.88 &  0.70 \\
21506 & 4.212 & 0 & 9.885 &  70.42 &  9.14 & 0.230 & 8.522 & 1.80 &  2.62 &  3.50 \\
25875 & 4.379 & 0 & 9.625 &  44.38 & 10.57 & 0.140 & 8.158 & 0.56 &  0.20 &  2.79 \\
26960 & 4.444 & 0 & 9.590 &   6.27 &  1.61 & 0.040 & 8.686 & 0.50 &  0.20 &  0.12 \\
31211 & 4.530 & 0 & 9.143 &   1.55 &  1.11 & 0.020 & 8.833 & 0.28 &  1.58 &  0.12 \\
33501 & 4.591 & 0 & 9.493 &  12.02 &  3.88 & 0.190 & 8.685 & 1.45 &  0.61 &  3.94 \\
\ \\
\hline
Averages & z = 4 \\
\hline
All                    & 4.339 & & 9.772 & 15 & 3.00 & 0.12 & 8.67 & 0.86 & 2.1 & 2.94 \\
\hline
\ \\
{\bf z = 5} \\
Non-detections \\
  6741 & 5.477 & 0 & 9.837 &  41.58 &  6.03 & 0.250 & 8.545 & 0.89 &  3.46 &  1.30 \\
 7310 & 4.948 & 0 &10.090 &  81.49 &  6.79 & 0.180 & 8.449 & 0.35 &  4.21 &  0.83 \\
10595 & 4.961 & 0 &10.179 &  32.01 &  2.13 & 0.220 & 8.672 & 0.28 &  4.05 &  6.82 \\
12383 & 4.796 & 0 & 9.265 &   3.95 &  2.19 & 0.060 & 8.750 & 0.33 &  8.00 &  0.15 \\
17427 & 4.893 & 0 & 9.336 &  10.24 &  4.66 & 0.100 & 8.404 & 0.17 &  0.65 &  1.93 \\
19247 & 4.974 & 0 & 9.230 &   9.79 &  5.76 & 0.060 & 8.547 & 1.64 &  8.00 & 11.43 \\
21665 & 4.960 & 0 & 9.265 &  18.75 & 10.42 & 0.260 & 8.347 & 1.42 &  8.00 &  1.62 \\
23926 & 5.129 & 0 & 9.851 &  46.70 &  6.58 & 0.200 & 8.443 & 0.72 &  0.20 &  1.12 \\
32882 & 4.659 & 0 & 9.559 &   3.30 &  0.92 & 0.060 & 8.648 & 1.80 &  3.27 &  0.65 \\
33659 & 4.991 & 0 & 9.083 &   3.25 &  2.71 & 0.070 & 8.561 & 1.00 &  1.83 &  0.18 \\
\ \\
\hline
Averages & z = 5 \\
\hline
All                    & 4.979 & & 9.730 & 14 & 5.21 & 0.15 & 8.55 & 0.86 & 2.5 & 2.60 \\
\enddata
\tablenotetext{1}{No AGN detected = 0; AGN suspected based on x-ray, optical, IR or radio data = 1}
\tablenotetext{2}{Median values are given for the `average' SFR and sSFR.}
\end{deluxetable*}

\newpage

\begin{deluxetable*}{ccccccc}
\tablewidth{0pt}
\tabletypesize{\scriptsize}

\tablecaption{Summary of observational parameters\label{tab:observations}}

\tablehead{
\multicolumn{1}{c}{(1)}      &
\multicolumn{1}{c}{(2)}      &
\multicolumn{1}{c}{(3)}      &
\multicolumn{1}{c}{(4)}      &
\multicolumn{1}{c}{(5)}      &
\multicolumn{1}{c}{(6)}      &
\multicolumn{1}{c}{(7)}       \\
%
\multicolumn{1}{c}{Redshift bin}                        &
\multicolumn{1}{c}{ALMA band}                        &
\multicolumn{1}{c}{$\lambda_{rest}$}                &
\multicolumn{1}{c}{\# targets}                            &
\multicolumn{1}{c}{\# detections}                       & 
\multicolumn{1}{c}{Beam}                                  &
\multicolumn{1}{c}{rms Noise}                            \\
\multicolumn{1}{c}{}                              &
\multicolumn{1}{c}{}                              &
\multicolumn{1}{c}{$\mu$m}                 &
\multicolumn{1}{c}{}                              &
\multicolumn{1}{c}{}                              &
\multicolumn{1}{c}{arcsec}                   &
\multicolumn{1}{c}{$\mu$Jy/beam}      \\
}
\startdata
2 & 6 & 290 & 20 & 15 & 0.69$\times$0.48 & 87.9 \\
3 & 6 & 220 & 20 & 11 & 0.56$\times$0.40 & 90.7 \\
4 & 7 & 260 & 20 &  0 & 0.72$\times$0.61 & 35.2 \\
5 & 7 & 220 & 10 &  0 & 0.76$\times$0.65 & 25.1 \\
\enddata
\end{deluxetable*}

\begin{deluxetable*}{ccrccc}
\tablewidth{0pt}
\tabletypesize{\scriptsize}

\tablecaption{Derived values for stacked galaxies\label{tab:stacked}}

\tablehead{
\\
\multicolumn{1}{c}{(1)}      &
\multicolumn{1}{c}{(2)}      &
\multicolumn{1}{c}{(3)}      &
\multicolumn{1}{c}{(4)}      &
\multicolumn{1}{c}{(5)}      &
\multicolumn{1}{c}{(6)}      \\
\ \\
\multicolumn{1}{c}{z bin}                                                     &
\multicolumn{1}{c}{$\bar{z}$}                                              &
\multicolumn{1}{c}{N$_{\mathrm{gal}}$\tablenotemark{1}}                             &
\multicolumn{1}{c}{$\sigma$}                                              &
\multicolumn{1}{c}{$\log{M_{\mathrm{gas}}/M_{\odot}}$}   &
\multicolumn{1}{c}{$\bar{f}_{gas}$}                                    \\ 
\multicolumn{1}{c}{}                              &
\multicolumn{1}{c}{}                              &
\multicolumn{1}{c}{}                              &
\multicolumn{1}{c}{$\mu$Jy/beam}      &
\multicolumn{1}{c}{}                              &
\multicolumn{1}{c}{}                              \\
}
\startdata
2 & 2.128 &   5 & 37.9 & $<$9.830 & $<$0.08 \\
3 & 3.270 &   9 & 32.4 & $<$9.762 & $<$0.15 \\
4 & 4.339 & 20 &   8.6 & $<$9.560 & $<$0.38 \\
5 & 4.979 & 10 &   8.0 & $<$9.505 & $<$0.37 \\
\enddata
\tablenotetext{1}{$N_{\mathrm{gal}}$ is the number of stacked galaxies for each redshift bin.}
\tablecomments{The upper limits correspond to 3$\sigma$ values of the rms noise in the stacked images.}
\end{deluxetable*}

\begin{figure*}
\epsscale{1.0}
\plotone{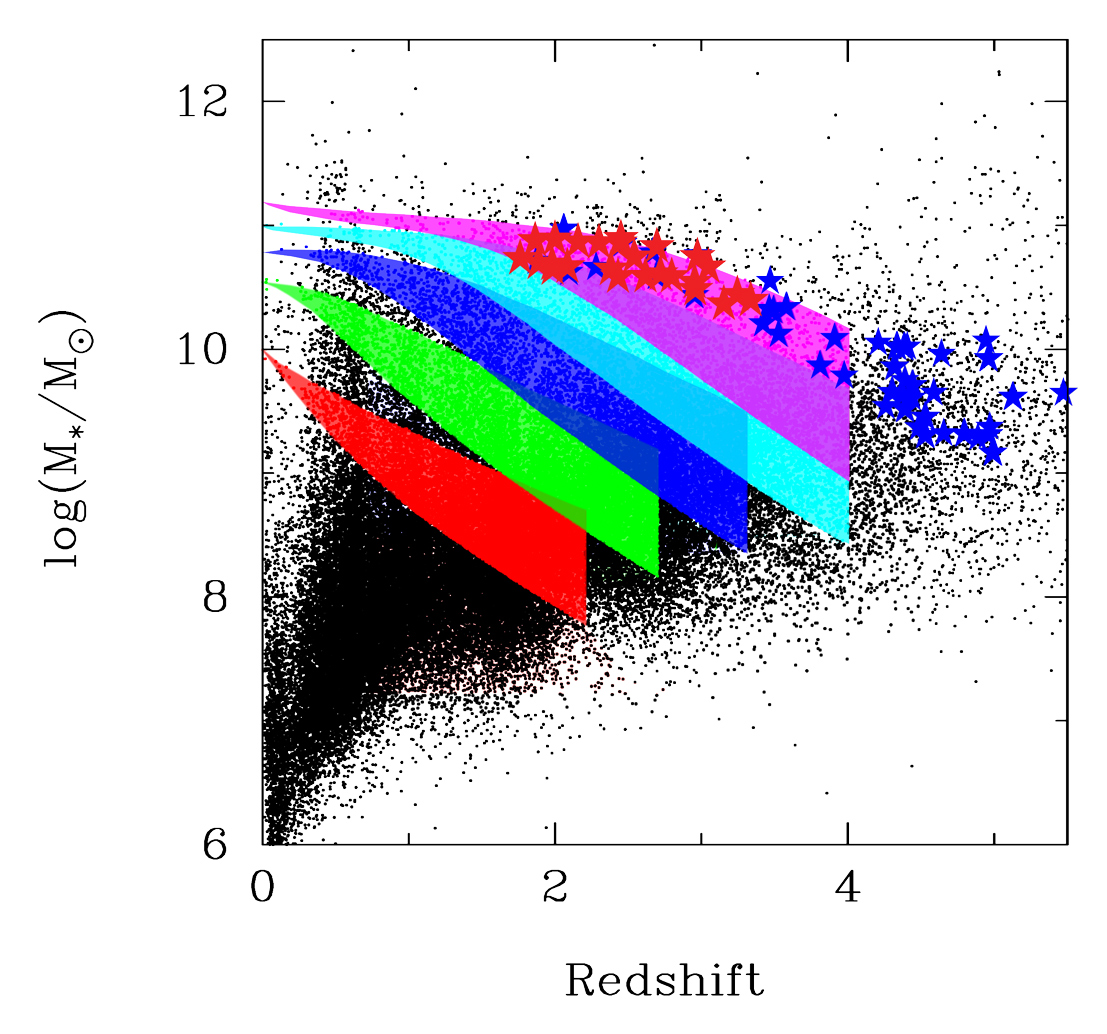}
\caption{Stellar mass versus redshift for $\sim$35,000 galaxies in the CANDELS GOODS-S field. The colored regions outline where progenitors
are located for $z$=0 galaxies of stellar mass $\log{(M_{*}/M_{\odot})}$ = 10.0, 10.5, 11.0 and 11.2 (red, green, blue, magenta and purple) using
the MEAM selection method (Moster et al. 2013; Behroozi et al. 2013). The spread of the selection slices represents the 1$\sigma$ uncertainty
as defined in Behroozi et al. (2013). The stars show the location of the 70 galaxies in our sample; red color indicates submm detected ones, and
blue undetected galaxies.}
\label{fig:slices}
\end{figure*}

\begin{figure*}
\epsscale{1.15}
\plotone{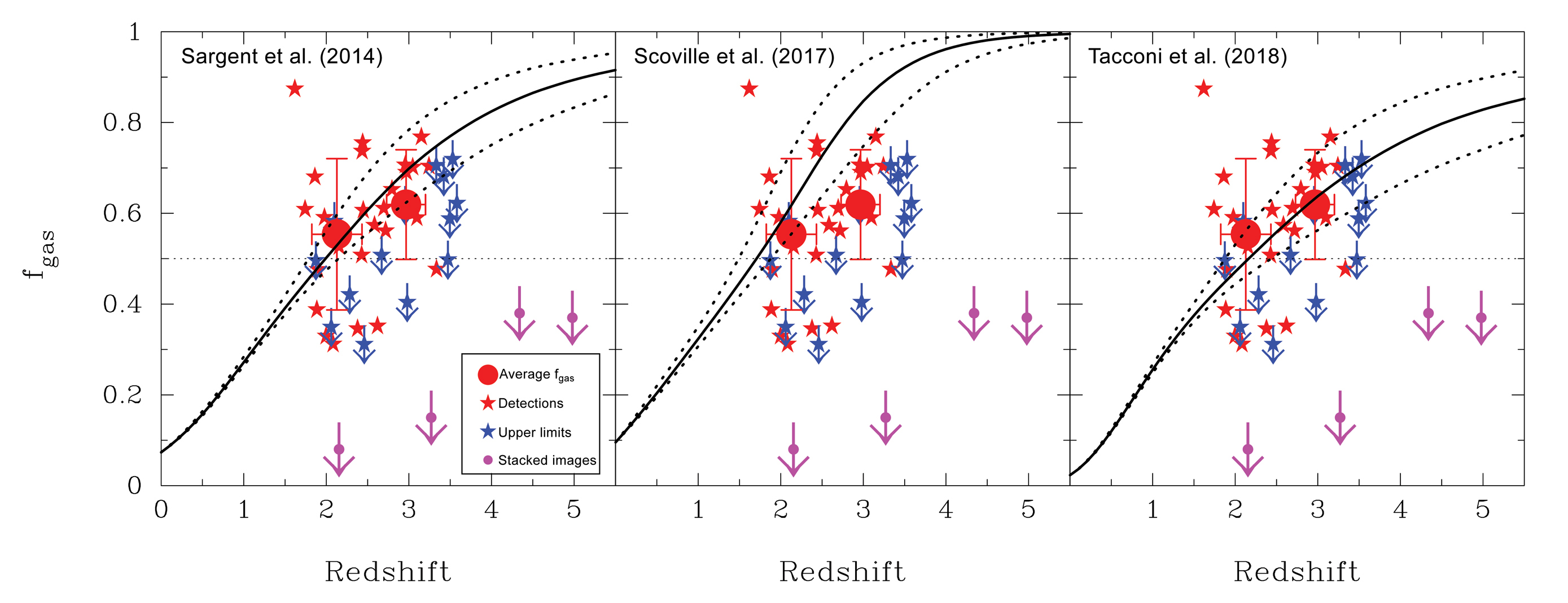}
\caption{Gas mass fraction vs. redshift for our sample: small red stars show individually detected galaxies, small blue stars indicate 3$\sigma$ upper limits
(only shown for the $z$=2 and 3 samples). The large red circles show the average gas mass fraction for the z=2 and z=3 samples. The purple circles show
the 3$\sigma$ upper limits to the gas mass fraction obtained from stacking of galaxies without detectable submm emission, including the entire $z$=4 and
$z$=5 samples. The gas mass fractions have been corrected for CMB effects (da Cunha et al. 2013). The curves show the expected gas mass fraction for
main sequence galaxies with a stellar mass equal to the MEAM selection criteria (Sect.~\ref{sec:meam}) derived from three different scaling relations
(Sargent et al. 2014; Scoville et al. 2017; Tacconi et al. 2018). The dashed lines correspond to the lower and upper limits of the MEAM stellar mass selection.
Apart from the upper limit to the stacked images, the observed gas mass fraction increases in accord with expectations from the scaling laws.}
\label{fig:scalinglaws}
\end{figure*}

\begin{figure*}
\epsscale{1.15}
\plotone{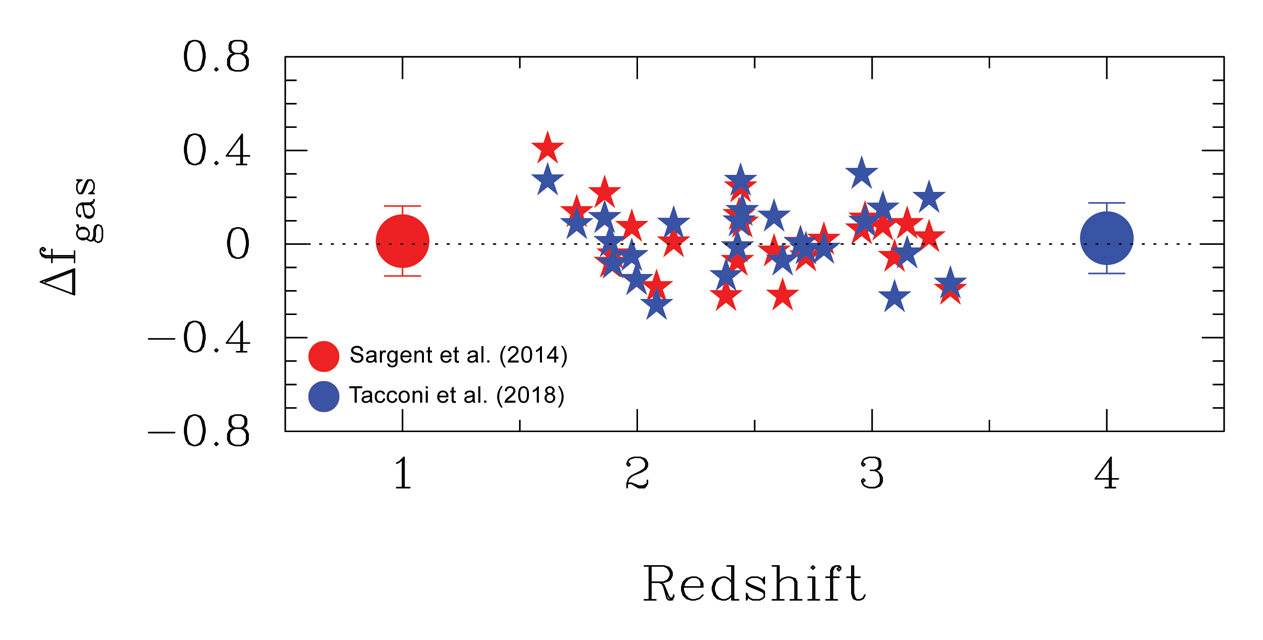}
\caption{The difference in gas mass fraction $\Delta f_{\mathrm{gas}}(z,M_*)$, the difference between the observationally derived gas fraction for individual
galaxies in our sample and that derived from the scaling laws of Sargent et al. (2014), red stars, and Tacconi et al. (2018), blue stars.
The red and blue circles represent averages of the two samples and the error bar shows the standard deviation of the mean. The Sargent et al. (2014) model
has an average $\Delta f_{\mathrm{gas}} = 0.013 \pm 0.149$, and the Tacconi et al. (2018) model has an average $\Delta f_{\mathrm{gas}} = 0.025 \pm 0.151$}
\label{fig:scatter_0}
\end{figure*}

\begin{figure*}
\epsscale{1.15}
\plotone{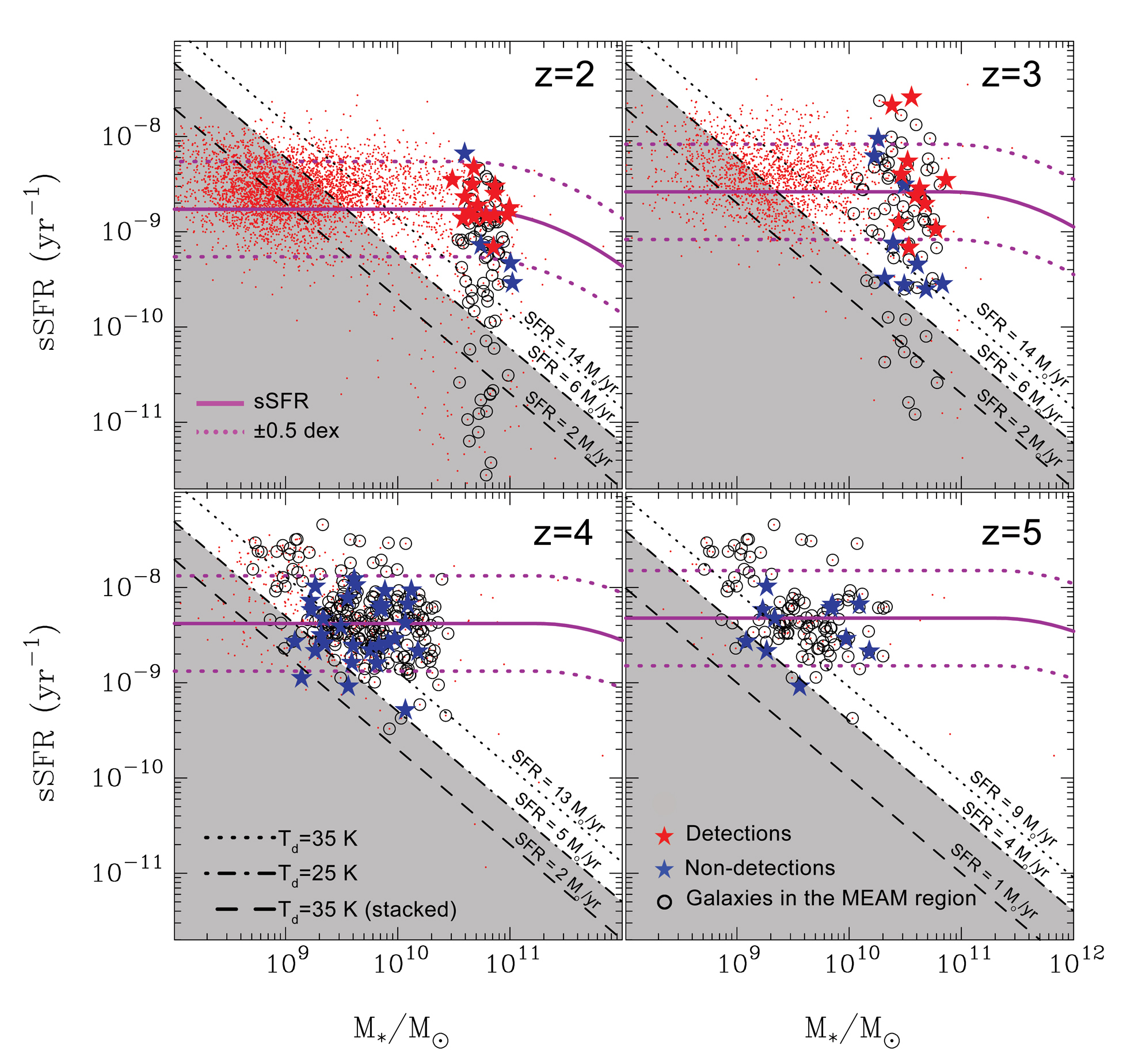}
\caption{
The specific star formation rate (sSFR) as a function of stellar mass for the four redshift bins making up our sample. The red dots correspond to all CANDELS
GOODS-S galaxies in the given redshift interval, the black circles correspond to galaxies located within our $M_*-z$ selection region. The red and blue stars
are the randomly selected galaxies observed with ALMA; red for submm deteced galaxies, blue for non-detections. The width of the stellar mass selection
becomes wider for higher redshift bins, illustrating the widening of the MEAM selection of potential progenitors (see Figure~\ref{fig:slices}).The diagonal lines
represent lower limits to the SFR for which submm emission would be detected with the present data sets given their 3$\sigma$ noise limits. The dash-dot line
corresponds to a dust SED with an assumed dust temperature of $T_{\mathrm{d}}=25$\,K, the dotted line to $T_{\mathrm {d}}=35$\,K. The dashed line is the limit
for the stacked images assuming $T_{\mathrm {d}}=35$\,K. Galaxies outside the shaded region should be detectable in the current ALMA data and shows that most
galaxies, even in the $z\geq4$ redshift bins fall in the detectable, non-shaded, region. The purple line is the specific star formation rate for the specific redshift
characterizing each redshift bin (from Schreiber et al. 2015). The dotted purple line is $\pm$0.5 dex of the sSFR.}
\label{fig:ssfr}
\end{figure*}

\begin{figure*}
\epsscale{1.15}
\plotone{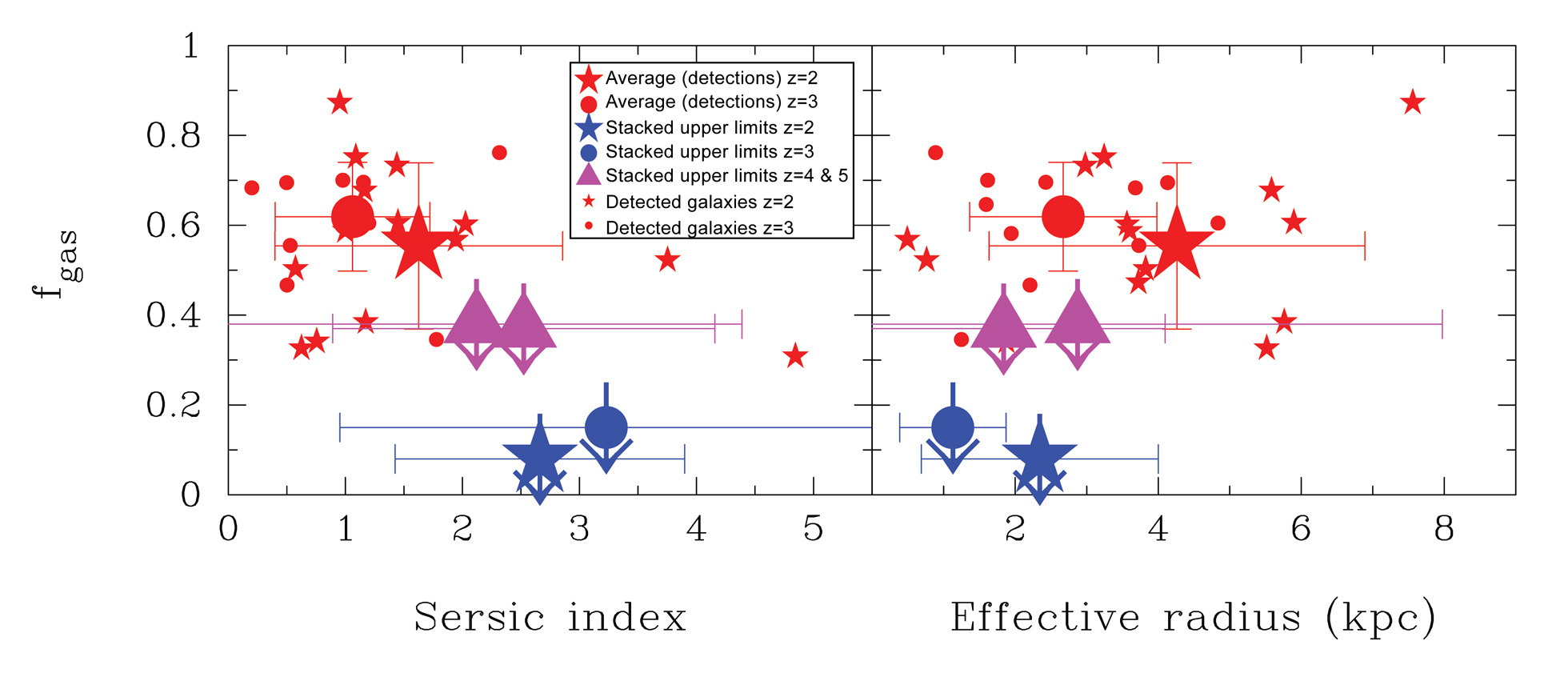}
\caption{The gas mass fraction versus S\'{e}rsic index, $S$ (left panel) and effective radius, $R_{\mathrm{E}}$ (right panel). Individual galaxies with
detected submm emission are shown as small red stars ($z$=2 sample) and small red circles ($z$=3 sample). The average values and the dispersion
of the mean for the detected galaxies are shown as a large red star and large red circle, respectively. The 3$\sigma$ upper limits of the stacked undetected
$z$=2 and $z$=3 galaxies are shown as blue star and circle. Likewise, the 3$\sigma$ upper limits to the stacked $z$=4 and 5 galaxies are shown as purple
triangles. Galaxies with $S\leq2.5$ are considered disk-dominated and those with $S>2.5$ are considered as spheroidal systems.
On average, galaxies with detected dust emission are disk-like while those without detectable emission tend to be spheroidal. Furthermore, the detected galaxies
have larger effective radii than those without detectable dust emission.} 
\label{fig:scatter_1}
\end{figure*}

\begin{figure*}
\epsscale{1.15}
\plotone{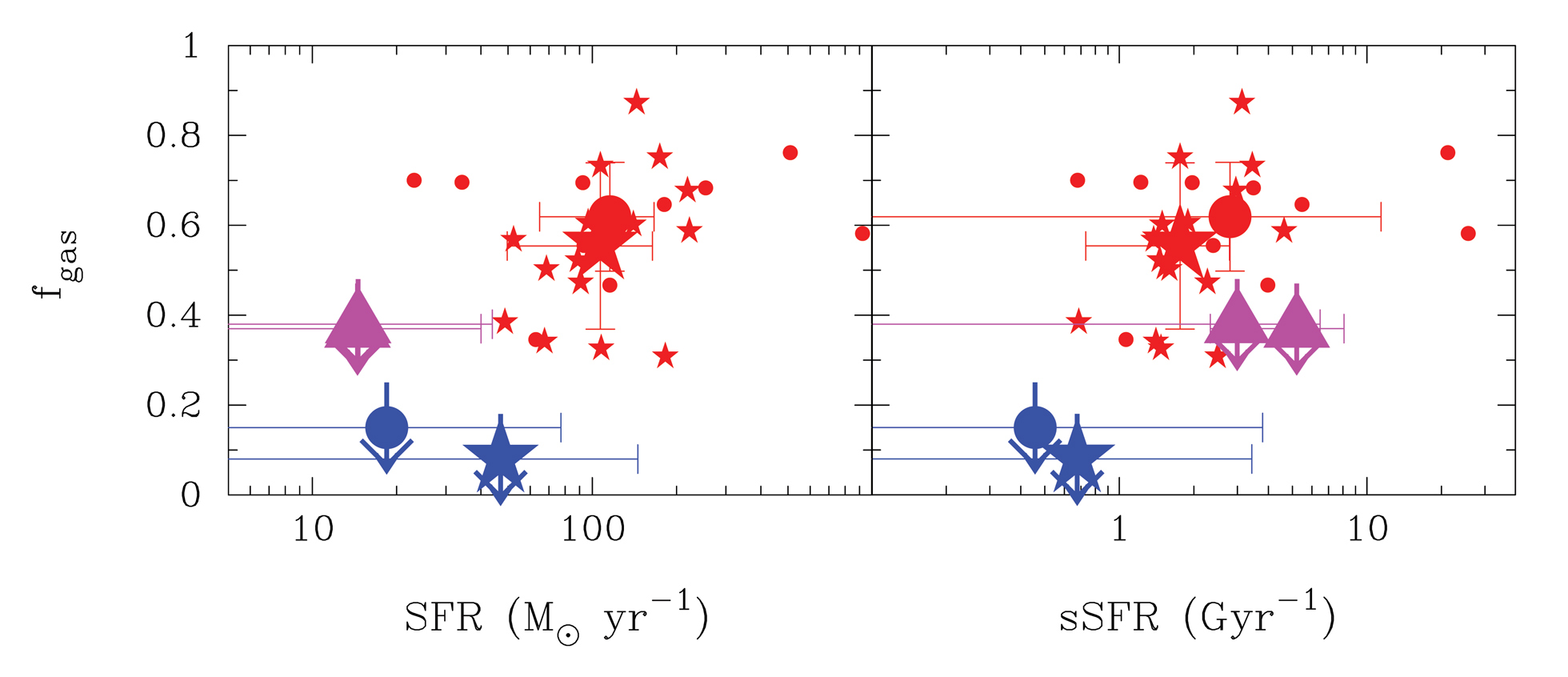}
\caption{The gas mass fraction versus star formation rate, SFR (left panel) and the specific star formation rate, sSFR (right panel). Designations and markings
are the same as in Figure~\ref{fig:scatter_1}. Here we plot the median SFR and sSFR values in logarithmic scale, together with the dispersion of the mean.
The large dispersion for the $z$=3 sSFR value from two galaxies (CANDELS ID\#4438 and 4878) with very high sSFR values; 26 and 21\,Gyr$^{-1}$, respectively). 
Galaxies with detected dust emission have, on average, higher SFRs than those without detectable dust emission. Still the average SFRs for our undetected
galaxies are not zero; the $z\sim2-3$ undetected galaxies have SFRs $\sim20-50$ M$_{\odot}$\,yr$^{-1}$ while the $z$=4 and 5 samples have SFRs $\sim$10\,M$_{\odot}$\,yr$^{-1}$.
This trend is reflected for the specific star formation rates, except for the $z\gtrsim4$ galaxies, which have higher sSFRs comparable to the $z\sim2-3$ detected galaxies.} 
\label{fig:scatter_2}
\end{figure*}

\begin{figure*}
\epsscale{1.15}
\plotone{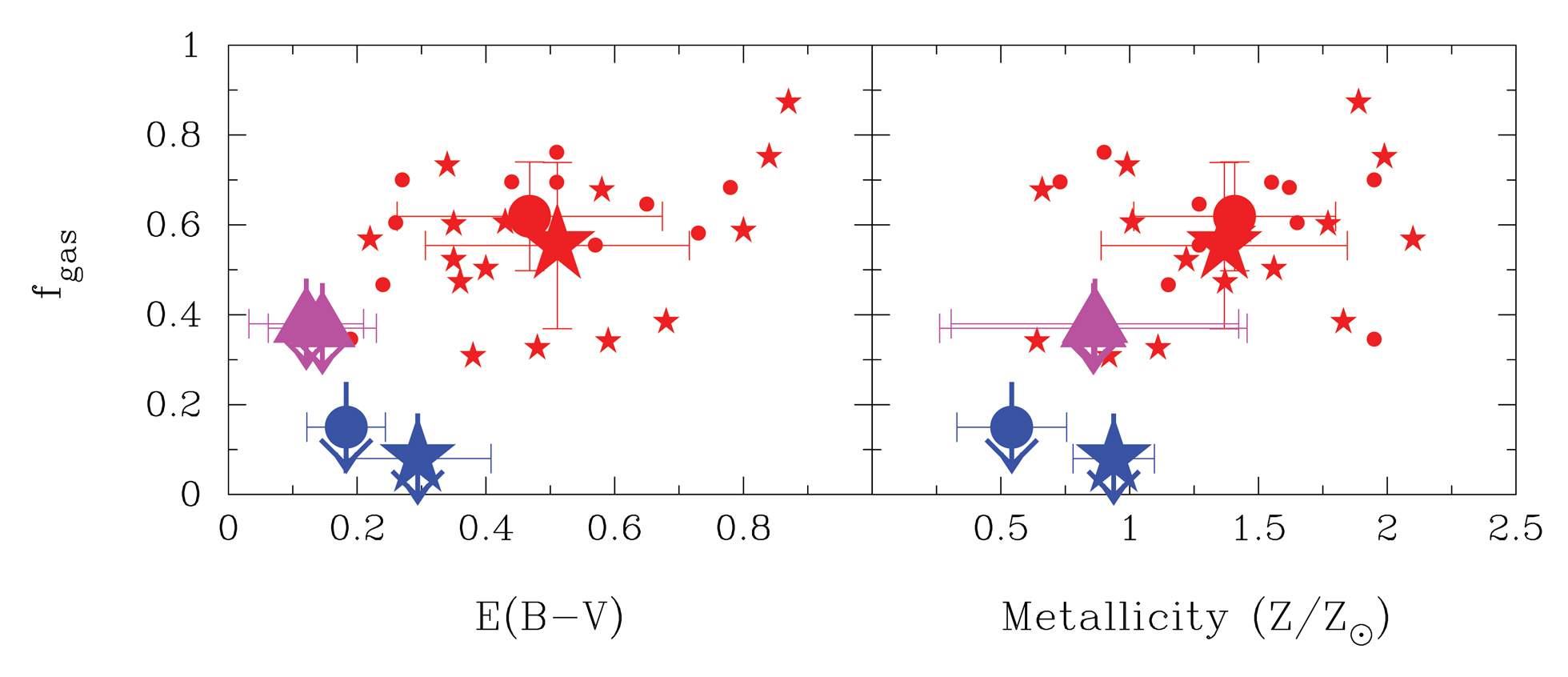}
\caption{The gas mass fraction versus the dust extinction, $E_{\mathrm{B-V}}$ (left panel) and stellar metallicity (right panel). Designations and markings
are the same as in Figure~\ref{fig:scatter_1}.  Not surprisingly, galaxies with detectable submm emission have, on average, higher extinction values
than those that are undetected. The stellar metallicities are derived from SED fits and show the same trend as the extinction values. The $z\gtrsim4$
galaxies (purple triangles) exhibit surprisingly high stellar metallicities, well in excess of the predicted gas-phase metallicity. The SED based metallicities
are discussed in Sec.~\ref{sec:metallicity}.}
\label{fig:scatter_3}
\end{figure*}

\begin{figure*}
\epsscale{1.15}
\plotone{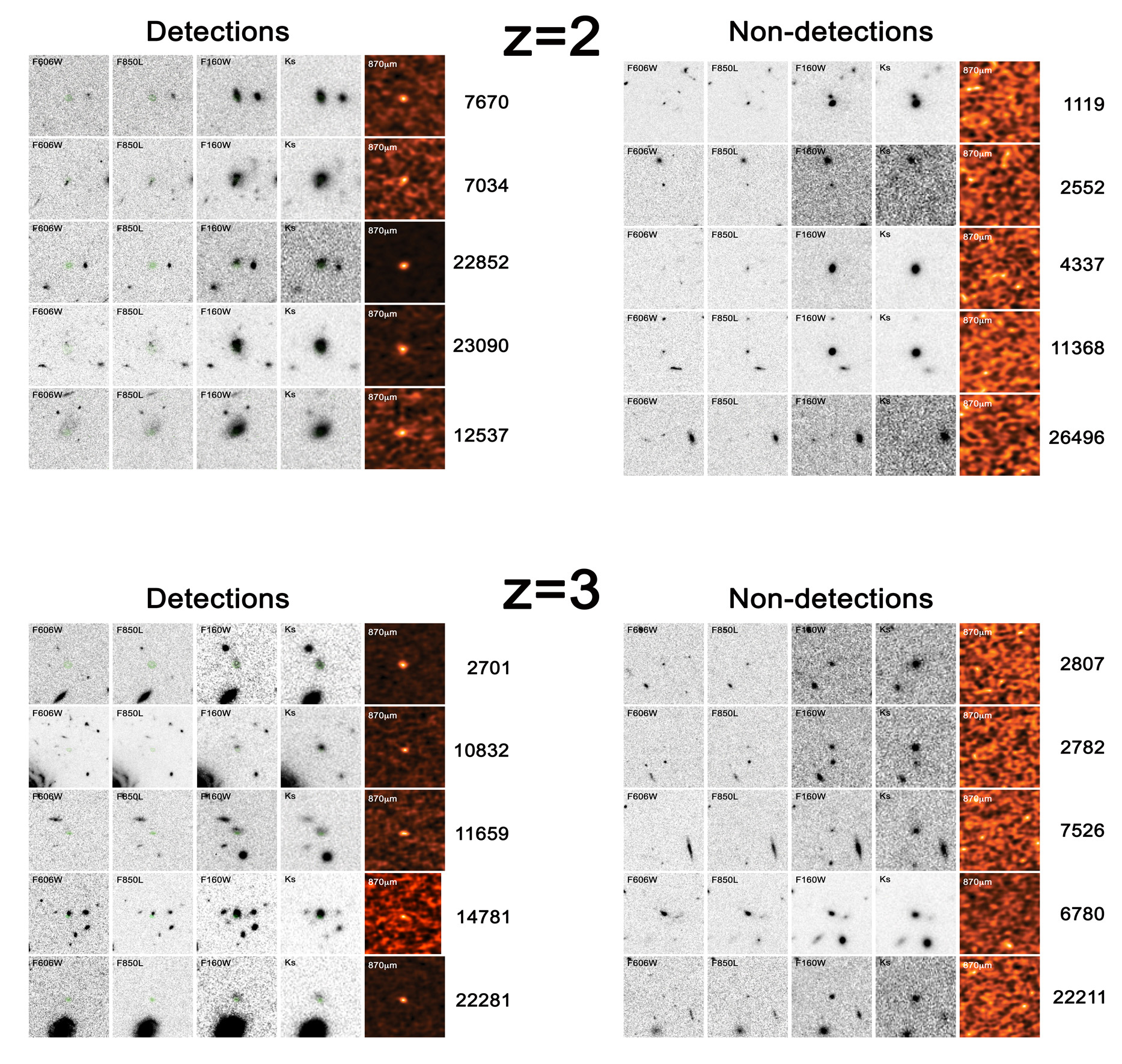}
\caption{Examples of CANDELS HST images and ALMA 870$\mu$m images of five detected and non-detected sources for the $z$=2 and $z$=3 samples.
The images have been selected to show the diversity of the target galaxies as well as its surroundings.
Each image is 7\ffas0$\times$7\ffas0 across. The images shown are, in order from left to right: HST/ACS F606W, HST/ACS F850LP, HST/WFC3 F160W,
VLT/Hawk-I Ks and ALMA 870$\mu$m. The ALMA images are not primary beam corrected and the stretch has been optimized to show the dust continuum
emission.}
\label{fig:sources_1}
\end{figure*}

\begin{figure*}
\epsscale{1.15}
\plotone{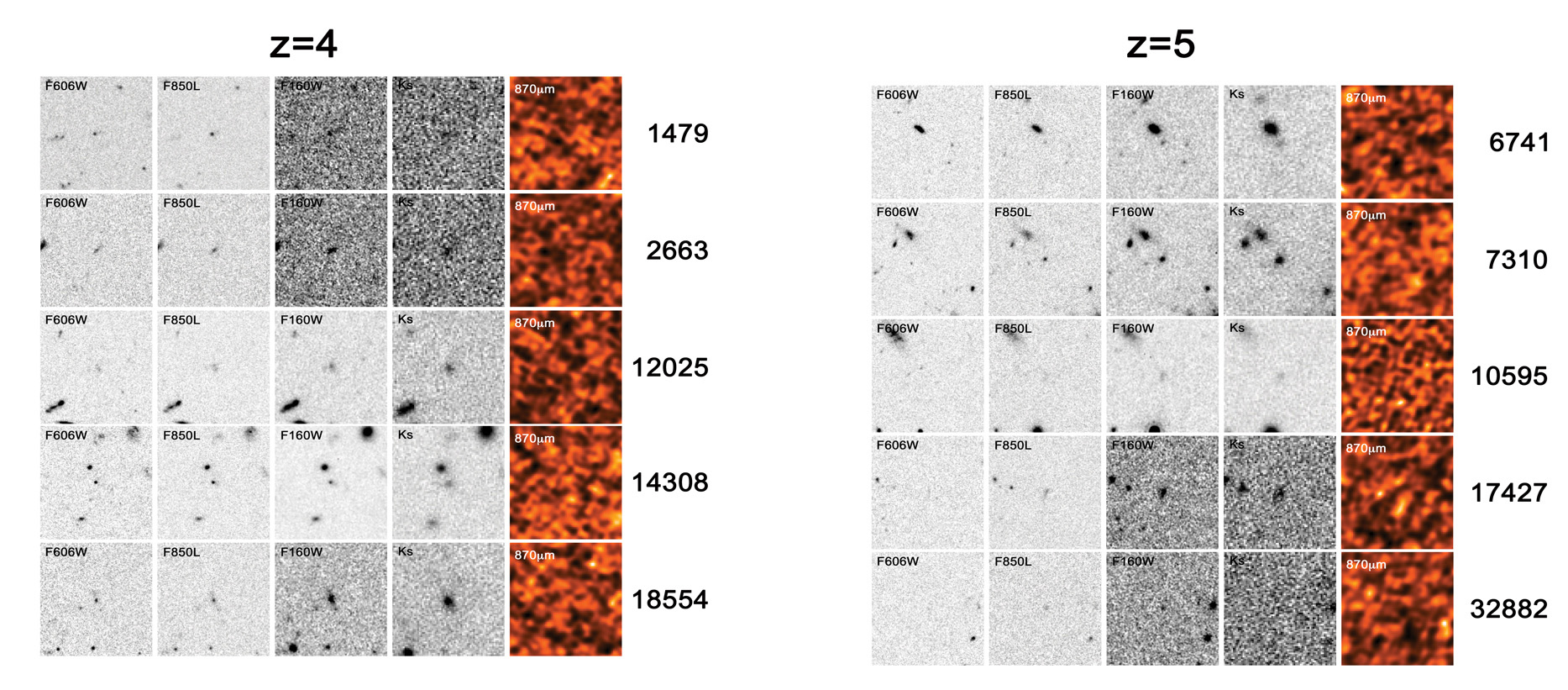}
\caption{Examples of CANDELS HST images and ALMA 1300$\mu$m images of five galaxies from the $z$=4 and $z$=5 samples. None of these galaxies are
submm detected. Each image is 7\ffas0$\times$7\ffas0 across. The images shown are, in order from left to right: HST/ACS F606W, HST/ACS F850LP, HST/WFC3
F160W, VLT/Hawk-I Ks and ALMA 870$\mu$m. The ALMA images are not primary beam corrected.}
\label{fig:sources_2}
\end{figure*}

\begin{figure*}
\epsscale{1.2}
\plotone{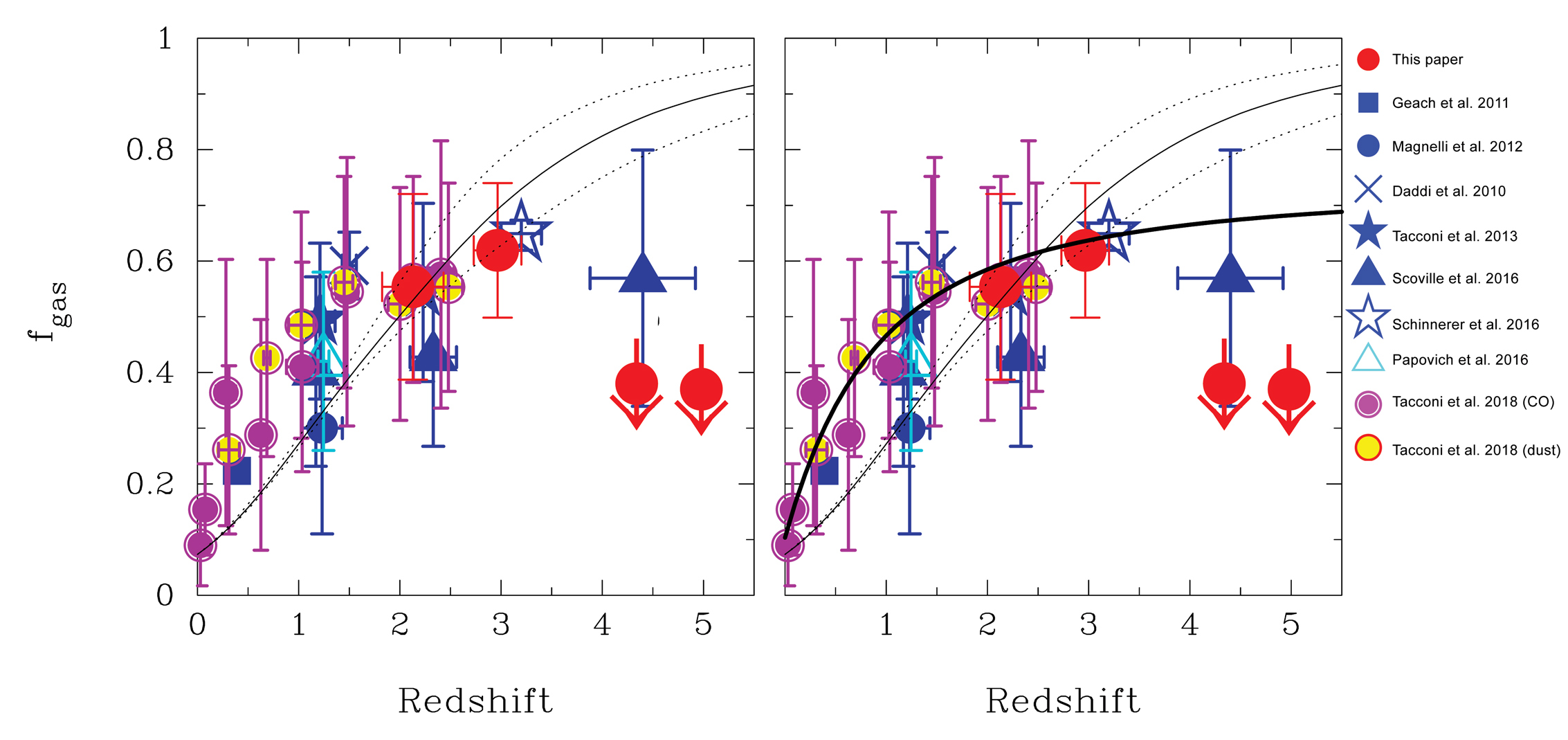}
\caption{The gas mass fraction versus stellar mass for this study and others compiled from the literature. Scoville et al. (2016), Schinnerer et al. (2016), part of
Tacconi et al. (2018) and this study use dust continuum as a proxy for the ISM gas mass. The others use CO line transitions, ranging from the J=1-0 to J=3-2
transition. The selection of the different samples all apply different criteria and may not be directly comparable to each other or our sample. Nevertheless, a
clear increase of the gas mass fraction is seen with increasing redshifts. The thin black line shows the expected gas mass fraction for main sequence
galaxies with a stellar mass equal to the MEAM selection criteria (Sect.~\ref{sec:meam}) derived from the scaling relation of Sargent et al. (2014). The dashed
lines correspond to the lower and upper limits of the stellar mass selection.
In the right panel, the thick black line shows a fit of a linear decrease of the gas mass fraction with cosmic time from $z$=3 to $z$=0. The curve corresponds
to a constant decrease of the $f_{\mathrm{gas}}$ by $0.043\pm0.007$\,Gyr. The apparent downturn of the gas mass
fraction at $z\gtrsim4$ is at least partially due to the stellar mass difference used in calculating the expected $f_{\mathrm{gas}}$ and the average stellar mass
of the Scoville et al. (2016) sample (see Figure~\ref{fig:scoville}).}
\label{fig:averages}
\end{figure*}

\begin{figure*}
\epsscale{0.9}
\plotone{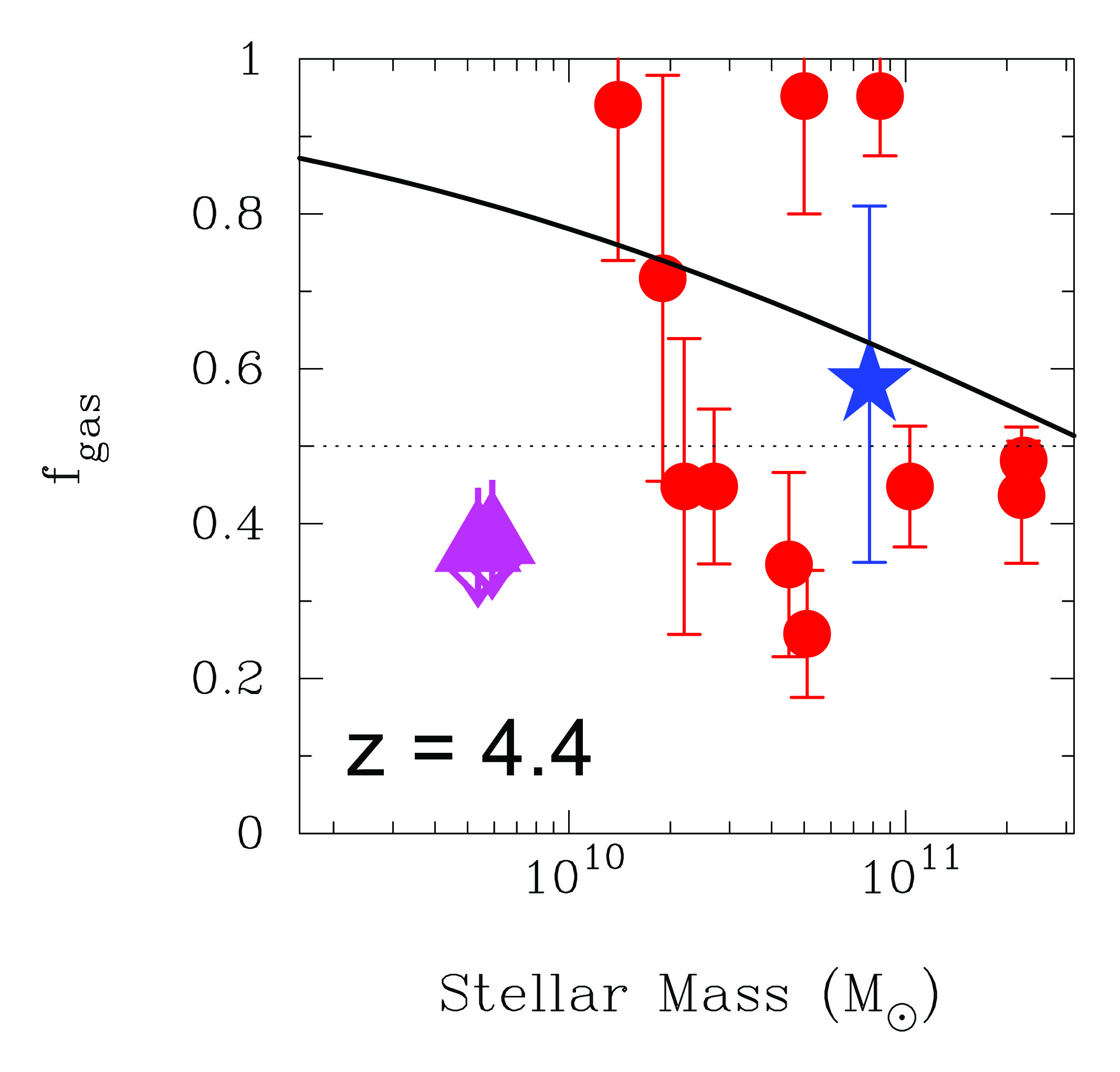}
\caption{The gas mass fraction versus stellar mass for our $z$=4 $z$=5 3$\sigma$ upper limits (purple triangles) and the $z$=4.4 sample from
Scoville et al. (2016) corrected for the CMB temperature at $z$=4.4 (red circles). The blue star is the average of the Scoville et al. values.
The line corresponds to the scaling relations from Sargent et al.
(2014) for a fixed redshift $z$=4.4.}
\label{fig:scoville}
\end{figure*}

\end{document}